%% file: main.tex
\ifdefined\pdfobjcompresslevel
\fi
\documentclass{article}

\let\preprintaddcontentsline\addcontentsline
\usepackage{iclr2027_conference,times}
\let\addcontentsline\preprintaddcontentsline
\iclrfinalcopy

\fancypagestyle{preprint}{%
  \fancyhf{}
  \fancyfoot[C]{\thepage}
  \renewcommand{\headrulewidth}{0.3pt}
  \renewcommand{\footrulewidth}{0pt}
}
\makeatletter
\renewcommand{\@maketitle}{%
  \begingroup
    \raggedright
    {\LARGE\scshape\hyphenpenalty=10000 \@title\par}
    \vspace{12pt}
    {\normalsize \@author\par}
  \endgroup
  \vspace{7pt}
}
\makeatother

\usepackage[utf8]{inputenc}
\usepackage[T1]{fontenc}
\usepackage{url}
\usepackage{booktabs}
\usepackage{amsfonts}
\usepackage{nicefrac}
\usepackage{color}
\usepackage[table]{xcolor}
\usepackage{colortbl}
\usepackage{pgf}
\usepackage{enumitem}
\usepackage{listings}
\usepackage{graphicx}
\usepackage{tabularx}
\usepackage{array}
\usepackage{pifont}
\usepackage{multirow}
\usepackage{wrapfig}
\usepackage{capt-of}
\usepackage{float}
\usepackage{needspace}
\usepackage{placeins}
\usepackage[most]{tcolorbox}
\usepackage{tikz}

\lstdefinestyle{mystyle}{
	backgroundcolor=\color{backcolour},
	commentstyle=\color{codegreen},
	language={Java}, 
	basicstyle=\sffamily\small,
	keywordstyle=\sffamily\small\bfseries\color{black},
	keywordstyle = [2]{\small\bfseries\color{teal}},
	keywordstyle = [3]{\small\bfseries\color{black}},
	keywordstyle = [4]{\small\bfseries\color{black}},
	otherkeywords = {assert},
	morekeywords = [2]{},
	morekeywords = [3]{},
	morekeywords = [4]{},
	breakatwhitespace=false,           
	captionpos=b,
	numbers=none,
	showspaces=false,                
	showstringspaces=false,
	showtabs=false, 
	tabsize=2
}

\lstdefinestyle{mystyle2}{
	backgroundcolor=\color{backcolour},
	commentstyle=\color{black},
	language={lisp}, 
	basicstyle=\ttfamily\scriptsize,
	keywordstyle=\scriptsize\bfseries\color{black},
	keywordstyle = [2]{\scriptsize\bfseries\color{teal}},
	keywordstyle = [3]{\scriptsize\bfseries\color{black}},
	keywordstyle = [4]{\scriptsize\bfseries\color{black}},
	keywordstyle = [5]{\scriptsize\color{black}},
	otherkeywords = {if, else, assert},
	morekeywords = [2]{ArrayList, LinkedHashMap, HashSet, Queue, LinkedList, TreeSet, Dictionary, Hashtable, TreeMap, LinkedHashSet, List, Map, Set, HashMap, HttpSession, Stack, ServletContext},
	morekeywords = [3]{put, add, get, contains, push, peek, addElement, pop, getAttribute, setAttribute,get, offer},
	morekeywords = [4]{get},
	morekeywords = [5]{;},
	breakatwhitespace=false,           
	captionpos=b,
	numbers=none,
	showspaces=false,                
	showstringspaces=false,
	showtabs=false, 
	tabsize=2
}

\newcommand{\mybox}[1]{
	\begin{tcolorbox}[
		boxsep=-0.5pt,
		standard jigsaw,
		boxrule=0.5 pt,
		opacityback=0,
		sharp corners
		]
		\linespread{1.2}\selectfont #1
	\end{tcolorbox}
}

\usetikzlibrary{positioning, arrows.meta, shapes.geometric, fit, backgrounds}

\definecolor{darkblue}{rgb}{0, 0, 0.5}
\definecolor{my_lightblue}{RGB}{194, 213, 247}
\definecolor{codebg}{RGB}{248,249,251}
\definecolor{codeframe}{RGB}{220,224,230}
\definecolor{codekeyword}{RGB}{31,78,121}
\definecolor{codecomment}{RGB}{92,99,112}
\definecolor{codestring}{RGB}{163,82,42}
\definecolor{linkblue}{HTML}{1A0DAB}

\usepackage[normalem]{ulem}
\usepackage{hyperref}
\hypersetup{
  colorlinks=true,
  allcolors=linkblue,
  pdfborder={0 0 0},
  pdfborderstyle={},
  pdfhighlight=/N,
  bookmarksopen=true,
  bookmarksopenlevel=0,
  bookmarksnumbered=true,
  bookmarksdepth=2,
  pdftitle={WitnessGym: Benchmarking Coding Agents on the Construction of Bug Witnesses},
  pdfauthor={Haomin QI, Xiangzhe Xu, Yiming Huang, Jingbo Shang, Chengpeng Wang},
  pdfsubject={Executable bug validation and automated benchmark construction}
}

\makeatletter
\def\NAT@hyper@#1{%
  \begingroup
  \let\hyper@natlinkbreak\@firstoftwo
  \protected@edef\WG@citation{#1}%
  \hyper@natlinkstart{\@citeb\@extra@b@citeb}%
  \expandafter\uline\expandafter{\WG@citation}%
  \hyper@natlinkend
  \endgroup
}
\let\WG@hyperlink\hyper@link
\def\hyper@link#1#2#3{\WG@hyperlink{#1}{#2}{\uline{#3}}}
\ExplSyntaxOn
\cs_new_protected:Npn \WitnessUrl #1
  {
    \tl_set:Nn \l_tmpa_tl {#1}
    \regex_replace_all:nnN { ([/._?=&\-]) } { \1 \c{penalty} 100 \c{relax} } \l_tmpa_tl
    \exp_args:NV \uline \l_tmpa_tl
  }
\ExplSyntaxOff
\let\WG@UrlFormatString\Url@FormatString
\def\Url@FormatString{\UrlFont\expandafter\WitnessUrl\expandafter{\Url@String}}
\makeatother
\providecommand{\doi}[1]{doi: \href{https://doi.org/#1}{\begingroup\urlstyle{rm}\nolinkurl{#1}\endgroup}}

\lstdefinestyle{auditjava}{%
  language=Java,
  basicstyle=\ttfamily\footnotesize,
  keywordstyle=\bfseries\color{codekeyword},
  commentstyle=\itshape\color{codecomment},
  stringstyle=\color{codestring},
  numberstyle=\scriptsize\color{gray},
  numbers=left,
  numbersep=8pt,
  showstringspaces=false,
  breaklines=true,
  breakatwhitespace=false,
  tabsize=2,
  frame=single,
  rulecolor=\color{codeframe},
  backgroundcolor=\color{codebg},
  xleftmargin=10pt,
  xrightmargin=6pt,
  framexleftmargin=6pt,
  framexrightmargin=4pt,
  aboveskip=4pt,
  belowskip=0pt,
  columns=fullflexible,
  keepspaces=true
}

\lstdefinestyle{appendixprompt}{%
  language={},
  basicstyle=\ttfamily\footnotesize\color{black},
  keywordstyle=\color{black},
  keywordstyle=[2]\color{black},
  keywordstyle=[3]\color{black},
  keywordstyle=[4]\color{black},
  keywordstyle=[5]\color{black},
  commentstyle=\color{black},
  stringstyle=\color{black},
  identifierstyle=\color{black},
  showstringspaces=false,
  breaklines=true,
  breakatwhitespace=false,
  tabsize=2,
  frame=single,
  rulecolor=\color{codeframe},
  backgroundcolor=\color{codebg},
  xleftmargin=10pt,
  xrightmargin=6pt,
  framexleftmargin=6pt,
  framexrightmargin=4pt,
  aboveskip=4pt,
  belowskip=0pt,
  columns=fullflexible,
  keepspaces=true,
  numbers=none
}

\newcommand{\toolname}{\textsc{WitnessGym}}

\newcolumntype{Y}{>{\raggedright\arraybackslash}X}
\newcolumntype{C}[1]{>{\centering\arraybackslash}p{#1}}
\newcolumntype{P}[1]{>{\centering\arraybackslash}p{#1}}
\newcolumntype{M}[1]{>{\centering\arraybackslash}m{#1}}

\newcommand{\yesmark}{\textcolor{green!45!black}{\ding{51}}}
\newcommand{\nomark}{\textcolor{red!70!black}{\ding{55}}}

\newcommand{\divLow}{\cellcolor[HTML]{F4D6D2}\textcolor[HTML]{8A2D25}{Low}}
\newcommand{\divMed}{\cellcolor[HTML]{FFF1C6}\textcolor[HTML]{7A5A00}{Medium}}
\newcommand{\divHigh}{\cellcolor[HTML]{DCEFD8}\textcolor[HTML]{2F6B3F}{High}}
\newcommand{\divOurs}{\cellcolor[HTML]{CDEBE5}\textcolor[HTML]{145A52}{High}}

\renewcommand{\mybox}[1]{
	\begin{tcolorbox}[
		boxsep=-0.5pt,
		standard jigsaw,
		boxrule=0.6pt,
		opacityback=0,
		sharp corners]
		#1
	\end{tcolorbox}
}

\newcommand{\actormaxdelta}{20}

\newcommand{\deltacell}[3]{%
  \begingroup
  \pgfmathsetmacro{\s}{min(abs(#1)/\actormaxdelta,1)}%
  \ifdim #1pt < 0pt
    \pgfmathsetmacro{\r}{1}%
    \pgfmathsetmacro{\g}{0.98 - 0.22*\s}%
    \pgfmathsetmacro{\b}{0.92 - 0.45*\s}%
  \else
    \pgfmathsetmacro{\r}{0.94 - 0.20*\s}%
    \pgfmathsetmacro{\g}{0.98 - 0.18*\s}%
    \pgfmathsetmacro{\b}{1}%
  \fi
  \edef\temp{\noexpand\cellcolor[rgb]{\r,\g,\b}\noexpand\strut\unexpanded{#2\;(#3)}}%
  \temp
  \endgroup
}

\newcommand{\ratecell}[1]{%
  \begingroup
  \pgfmathsetmacro{\s}{min(max(#1/72,0),1)}%
  \pgfmathsetmacro{\r}{1.00}%
  \pgfmathsetmacro{\g}{0.96 - 0.34*\s}%
  \pgfmathsetmacro{\b}{0.90 - 0.72*\s}%
  \edef\temp{\noexpand\cellcolor[rgb]{\r,\g,\b}\noexpand\strut\unexpanded{#1\%}}%
  \temp
  \endgroup
}

\newcommand{\weakcell}[1]{\cellcolor{red!10}#1}

\newcommand{\ppmaxdelta}{36}

\newcommand{\ppdownmark}[1]{\raisebox{-0.55ex}{\scriptsize\textcolor{red!72!black}{$\downarrow$#1}}}
\newcommand{\ppupmark}[1]{\raisebox{-0.55ex}{\scriptsize\textcolor{blue!65!black}{$\uparrow$#1}}}

\newcommand{\ppdowncell}[2]{%
  \begingroup
  \pgfmathsetmacro{\s}{min(abs(#1)/\ppmaxdelta,1)}%
  \pgfmathsetmacro{\r}{1.00}%
  \pgfmathsetmacro{\g}{0.985 - 0.145*\s}%
  \pgfmathsetmacro{\b}{0.965 - 0.205*\s}%
  \edef\temp{\noexpand\cellcolor[rgb]{\r,\g,\b}\noexpand\strut\unexpanded{#2\,\ppdownmark{#1pp}}}%
  \temp
  \endgroup
}

\newcommand{\ppupcell}[2]{%
  \begingroup
  \pgfmathsetmacro{\s}{min(abs(#1)/\ppmaxdelta,1)}%
  \pgfmathsetmacro{\r}{0.94 - 0.10*\s}%
  \pgfmathsetmacro{\g}{0.975 - 0.05*\s}%
  \pgfmathsetmacro{\b}{1.00}%
  \edef\temp{\noexpand\cellcolor[rgb]{\r,\g,\b}\noexpand\strut\unexpanded{#2\,\ppupmark{#1pp}}}%
  \temp
  \endgroup
}

\title{WitnessGym: Benchmarking Coding Agents on the Construction of Bug Witnesses}

\author{%
  {\bfseries
    Haomin QI\textsuperscript{1}\quad
    Xiangzhe Xu\textsuperscript{2}\quad
    Yiming Huang\textsuperscript{1}\quad
    Jingbo Shang\textsuperscript{1}\quad
    Chengpeng Wang\textsuperscript{3}}\\[7pt]
  {\small
  \begin{tabular*}{\textwidth}{@{}c@{\extracolsep{\fill}}cc@{}}
    \raisebox{-2.25pt}{\includegraphics[height=11.5pt]{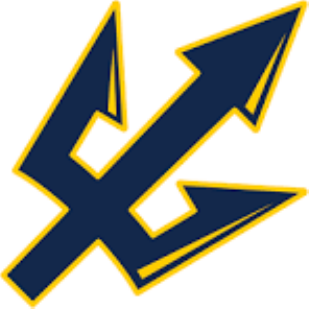}}\hspace{3pt}\textsuperscript{1}University of California, San Diego &
    \raisebox{-1.5pt}{\includegraphics[height=9pt]{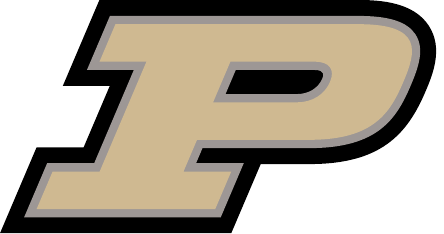}}\hspace{3pt}\textsuperscript{2}Purdue University &
    \raisebox{-2.25pt}{\includegraphics[height=11.5pt]{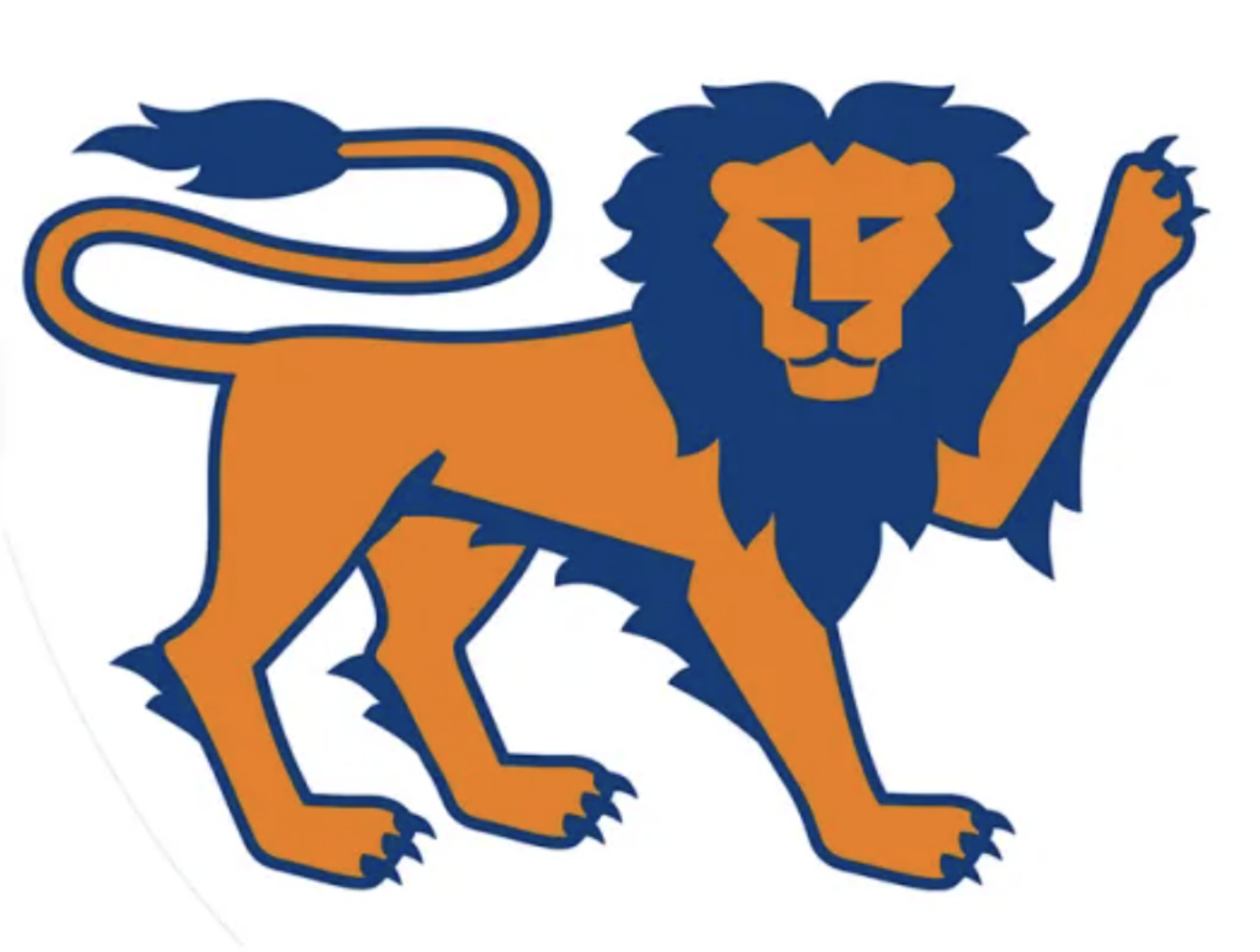}}\hspace{3pt}\textsuperscript{3}National University of Singapore
  \end{tabular*}}%
}

\begin{document}
\maketitle
\thispagestyle{preprint}

\input{sections/abstract}
\begingroup
\centering
\normalsize
\urlstyle{same}
\begin{tabular}{@{}c@{}}
  \raisebox{-2pt}{\includegraphics[height=10pt]{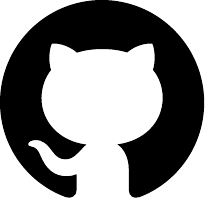}}\hspace{4pt}\href{https://github.com/HarminChee/WitnessGym}{https://github.com/HarminChee/WitnessGym} \\[3pt]
  \raisebox{-2pt}{\includegraphics[height=10pt]{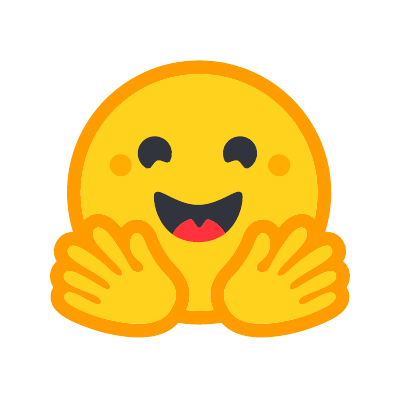}}\hspace{4pt}\href{https://huggingface.co/datasets/HarminChee/WitnessGym}{https://huggingface.co/datasets/HarminChee/WitnessGym}
\end{tabular}
\par
\vspace{0.4em}
\endgroup
\input{sections/introduction}
\input{sections/background}
\FloatBarrier
\input{sections/methodology}
\FloatBarrier
\input{sections/evaluation}
\FloatBarrier
\input{sections/related}
\input{sections/conclusion}

\begingroup
\hypersetup{allcolors=black}
\makeatletter
\let\hyper@link\WG@hyperlink
\let\Url@FormatString\WG@UrlFormatString
\makeatother
\bibliographystyle{iclr2027_conference}
\bibliography{reference}
\endgroup

\clearpage
\input{sections/appendix}

\end{document}

%% file: sections/abstract.tex
\begin{abstract}
\textbf{Bug validation} asks a coding agent to produce an \textbf{executable witness} for a reported bug. The witness combines a concrete input with a testing harness and exposes faulty behavior during execution. Such evidence makes audit findings actionable, yet benchmark evaluation is difficult when cases reuse public historical bugs and witnesses or require manual construction.
We present \textbf{\toolname{}}, an automated framework for constructing bug-validation benchmarks through bug injection. It injects bugs into test-reached paths of real projects, rebuilds each project, and retains cases exposed by a construction-time witness. Bug specifications and execution adapters allow extension to additional bug types and languages. Bug-preserving transformations vary the surrounding structure while preserving the witness behavior.
Based on real-world Java projects with test suites, \toolname{} automatically constructs 1{,}300 benchmark cases.
The injected patches resemble historical bug patches and are difficult for the two evaluated models to distinguish in blinded comparisons.
We evaluate four coding agent frameworks in six framework/model pairings across bug types, execution contexts, and transformation depths.
Witness construction remains difficult even when the bug pattern is known.
Our framework, benchmark cases, and evaluation scripts are available.
\end{abstract}

%% file: sections/introduction.tex
\section{Introduction}
Agentic coding systems increasingly address repository-level software-engineering tasks~\citep{wang2025openhands,yang2024swe,xia2025live,antoniades2024swe}, making reliable code auditing an important capability~\citep{guo2025repoaudit,wang2024llmdfa}.
In real-world AI coding scenarios, useful audit findings require concrete evidence that developers can inspect and act on, in addition to a suspicious code location~\citep{wang2026cybergym,zhang2025bountybench,zhu2025cvebench,cheng2025agentic,liu2025llm,DBLP:journals/corr/abs-2506-04962,DBLP:conf/acl/PengYDZZZGZ25}.
This requirement matters because bug reports generated by existing AI coding agents may still contain non-negligible false positives~\citep{zhao2026anypoc,du2025minimizing}, and repeated false alarms can quickly erode developers' trust in the tool.
For this reason, a useful coding agent should generate an \textbf{executable witness} that exposes the reported bug~\citep{wang2026cybergym,gezgin2026can}. An executable witness combines a concrete input with a testing harness that invokes the relevant code under the required environment and exposes an observable failure such as a crash or assertion violation. Such evidence helps developers distinguish real bugs from false alarms and decide whether a report deserves attention.

\begin{figure*}[t]
  \centering
  \includegraphics[width=\linewidth]{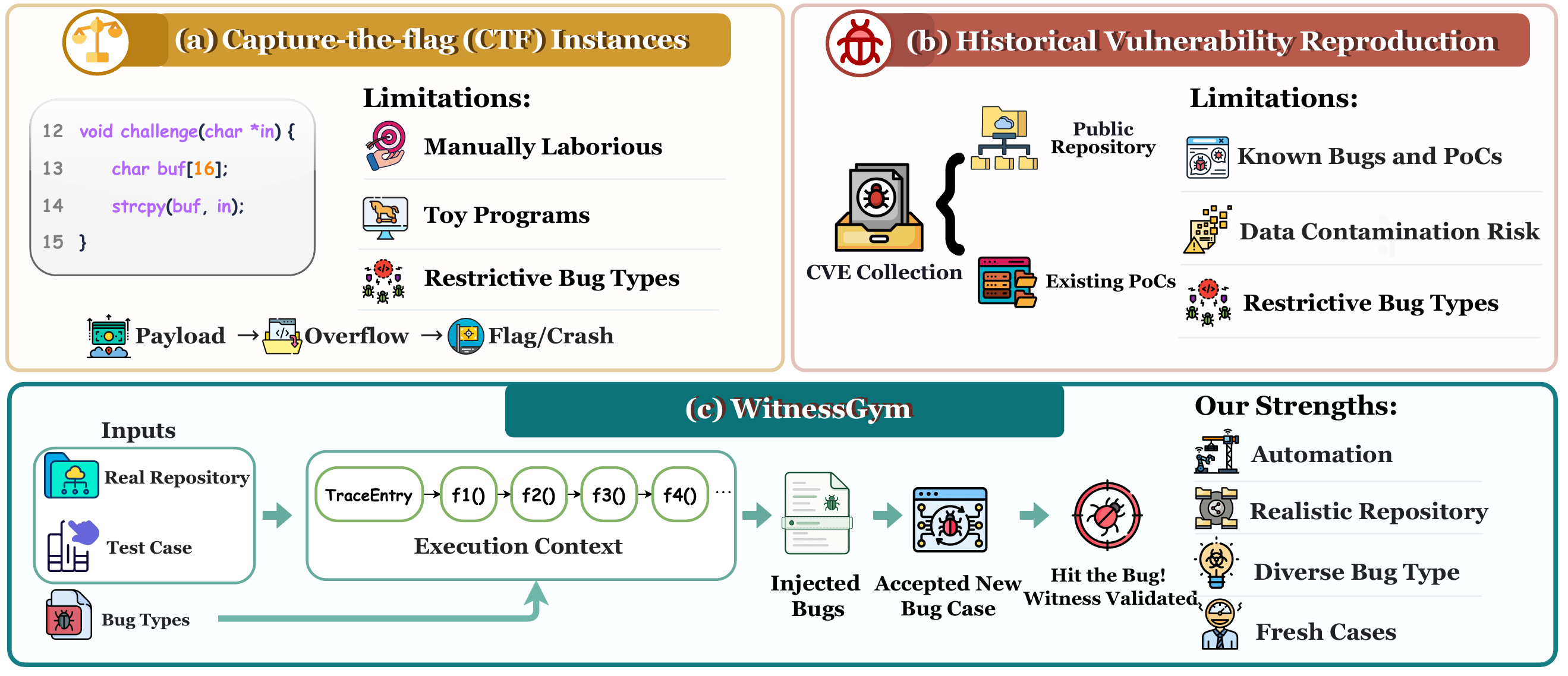}
  \caption{Comparison of benchmark-construction methodologies for executable bug validation.
  }
  \label{fig:background-motivation}
  \vspace{-10pt}
\end{figure*}

Existing benchmarks provide only partial support for evaluating bug validation on new cases in real project contexts. They typically rely on curated capture-the-flag (CTF) challenges or disclosed vulnerabilities, tying evaluation to manually assembled tasks or known bugs and witnesses (Figure~\ref{fig:background-motivation}a and b).
For example, NYU CTF Bench~\citep{shao2024nyuctfbench} and Cybench~\citep{zhang2025cybench} offer executable objectives in challenge-specific environments whose construction requires curation. Benchmarks based on CVEs~\citep{zhu2025cvebench}, bug-bounty reports~\citep{zhang2025bountybench}, and fuzzing artifacts~\citep{wang2026cybergym} retain real project context but draw their cases from historical vulnerabilities. Reusing these public artifacts limits the supply of new evaluation targets and creates opportunities for prior exposure to their solutions.

We target a benchmark that can \textbf{generate new cases automatically} as models evolve while \textbf{preserving the contextual reasoning} required in real code auditing~\citep{DBLP:conf/acl/WuPXZ0MD0LW25,pu2025dynamic,saxon2024benchmarks}. These goals address both the cost of case construction and the dependence on existing solutions. Even when a static-analysis query reports a recognizable bug pattern and its location, validation requires concrete object states, call sequences, and path conditions that expose the bug through an observable failure.

We introduce \toolname{}, a framework that meets these goals through automated bug injection. To construct new cases, it adapts API Contract, Value Flow, and Logic bug types from the CodeQL Java query library into injection specifications~\citep{codeql_java_queries}. These categories distinguish contract, data-state, and functional reasoning in witness construction (Section~\ref{sec:bug_injection}). To preserve real project context, it injects these bugs into production paths exercised by existing tests, rebuilds the project, and reruns the associated test. This test serves as a construction-time witness and must continue to expose the bug after transformations vary the surrounding call, data, and control structure. Its contents are removed before agent evaluation, leaving the agent to construct its own witness. Construction establishes that the bug is observable; evaluation measures whether an agent can independently recover the conditions needed to expose it.

We implement \toolname{} and automatically construct 1{,}300 benchmark cases using ten bug types in these three categories and six real-world Java projects with test suites, at a total cost of \$16{,}500.
We evaluate four agent frameworks instantiated in six framework/model pairings on their ability to construct executable witnesses for the benchmark cases.
Our evaluation shows substantial remaining headroom on this task.
The best-performing configuration, Codex with GPT-5.4, achieves 68.9\% average validation success. Performance varies across agents and models, with a 50-percentage-point gap between the highest and lowest averages. Value Flow and Logic bugs are generally harder than API Contract bugs, and success decreases as execution contexts lengthen or transformation depth increases. The evaluation costs approximately \$8{,}700.

This work makes three main contributions.
First, we present \toolname{}, a systematic methodology for \textbf{automatically constructing bug-validation benchmarks} by injecting diverse bug types into real-world projects, producing new bug--witness pairs.
Second, we instantiate this methodology using
ten bug types across API Contract, Value Flow, and Logic and six real-world Java projects, yielding \textbf{1{,}300 execution-validated cases}.
Third, we \textbf{evaluate coding agents} on the resulting benchmark and analyze the factors that affect bug-validation performance.

%% file: sections/background.tex
\section{Background and Motivation}
\label{sec:background}

\noindent\emph{\textbf{Bug Validation.}}
Given a repository, a bug description, and a reported location, bug validation asks an agent to construct an executable witness. The witness combines a concrete input with a testing harness that invokes the relevant code under the required environment and exposes the bug through an exception, assertion violation, crash, or another observable failure.

This task turns a reported bug into evidence that developers can reproduce and act on. A recognizable pattern or a localized report leaves the \textbf{concrete input, program state, and failure oracle} to be constructed. The agent must recover the relevant call sequence and path conditions, instantiate the required state, and expose the faulty behavior. For example, an API-contract violation may require a particular object state and call sequence before an assertion can expose the incorrect result. Evaluation therefore needs executable cases embedded in real program contexts, with a known failure that can serve as the validation oracle.

\noindent\emph{\textbf{Existing Benchmarks.}}
Table~\ref{tab:background-benchmark-comparison} compares the construction capabilities needed for this evaluation. Curated CTF tasks provide explicit objectives, but adding cases requires new challenges and associated environments~\citep{shao2024nyuctfbench,zhang2025cybench}. Historical bug-reproduction work, including LIBRO, Issue2Test, and CyberGym, retains real project context while relying on existing bug reports or disclosed vulnerabilities~\citep{libro2023,issue2test2026,wang2026cybergym}. Its case supply is consequently tied to the availability of historical artifacts. Bug-injection methods automate construction: LAVA and EvilCoder introduce new vulnerabilities through specialized injection mechanisms, while FixReverter applies patterns learned from past fixes at new injection sites~\citep{dolangavitt2016lava,pewny2016evilcoder,fixreverter2022}. Their construction objectives give limited control over the surrounding program structure for comparing agents across validation difficulties.

\begin{table}[H]
  \centering
  \small
  \setlength{\tabcolsep}{2.4pt}
  \renewcommand{\arraystretch}{1.06}
  \caption{Comparison of bug-validation benchmark construction. \textbf{Diversity} is high for multiple bug families, medium for variants within a dominant family, and low for one narrow family. \textbf{Auto. cases} are constructed automatically; \textbf{fresh bugs} are generated rather than reused; \textbf{real projects} retain production code with its build and test environments; and \textbf{context control} varies surrounding call, data, or control structure.}
  \label{tab:background-benchmark-comparison}
  {
  \begin{tabularx}{\textwidth}{@{}Y C{0.09\textwidth} C{0.11\textwidth} C{0.11\textwidth} C{0.13\textwidth} C{0.151\textwidth}@{}}
    \toprule
    \textbf{Work} & \textbf{Diversity} & \textbf{\mbox{Auto. cases}} & \textbf{\mbox{Fresh bugs}} & \textbf{\mbox{Real projects}} & \textbf{\mbox{Context control}} \\
    \midrule
    \multicolumn{6}{@{}l}{\textit{CTF and security challenges}} \\
    NYU CTF Bench~\citeyearpar{shao2024nyuctfbench} & \divHigh & \nomark & \nomark & \nomark & \nomark \\
    Cybench~\citeyearpar{zhang2025cybench} & \divHigh & \nomark & \nomark & \nomark & \nomark \\
    \addlinespace[1pt]
    \multicolumn{6}{@{}l}{\textit{Historical bug validation}} \\
    LIBRO~\citeyearpar{libro2023} & \divHigh & \nomark & \nomark & \yesmark & \nomark \\
    Issue2Test~\citeyearpar{issue2test2026} & \divHigh & \nomark & \nomark & \yesmark & \nomark \\
    CyberGym~\citeyearpar{wang2026cybergym} & \divMed & \nomark & \nomark & \yesmark & \nomark \\
    \addlinespace[1pt]
    \multicolumn{6}{@{}l}{\textit{Bug-injection methods}} \\
    LAVA~\citeyearpar{dolangavitt2016lava} & \divLow & \yesmark & \yesmark & \yesmark & \nomark \\
    EvilCoder~\citeyearpar{pewny2016evilcoder} & \divLow & \yesmark & \yesmark & \yesmark & \nomark \\
    FixReverter~\citeyearpar{fixreverter2022} & \divMed & \yesmark & \yesmark & \yesmark & \nomark \\
    \midrule
    \rowcolor{black!4}
    \toolname{} (Our Work) & \divOurs & \yesmark & \yesmark & \yesmark & \yesmark \\
    \bottomrule
  \end{tabularx}
  }
\end{table}

\noindent\emph{\textbf{Construction Goals.}}
We seek \textbf{automatic construction of new bug cases in real projects}, covering multiple bug families and retaining an executable oracle for every accepted case. Fresh cases reduce dependence on public bug--witness pairs, while \textbf{controlled structural variation} lets the benchmark probe how context affects validation. \toolname{} combines these properties by injecting bugs into test-reached code and applying transformations that preserve the observed failure. This yields new validation targets with realistic project dependencies and adjustable surrounding structure, without per-case manual bug design or witness authoring.

%% file: sections/methodology.tex
\section{Methodology}
\label{sec:methodology}

This section introduces \toolname{}, a framework for constructing bug-validation benchmarks. Given a real-world project with an existing test suite and a library of bug types, \toolname{} injects bugs into test-reached production contexts. The associated test provides the harness and concrete input used to expose each injected bug; we call this artifact the \textbf{construction-time witness}. Its contents are removed before agent evaluation.

\begin{figure}[ht]
\centering
\includegraphics[width=1\textwidth]{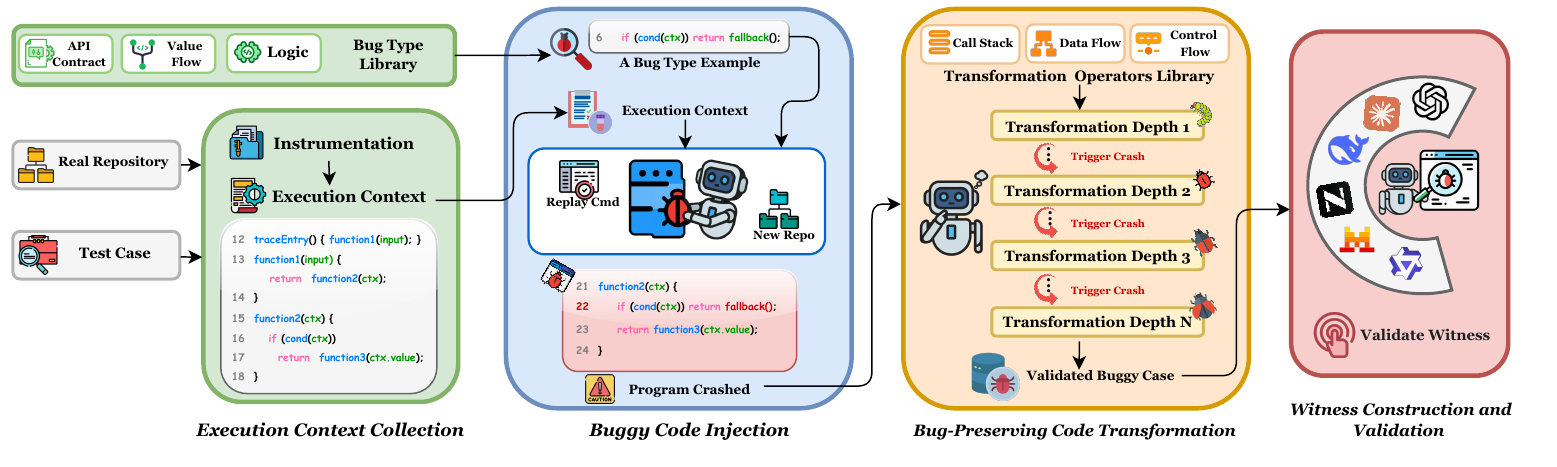}
\caption{Overview of the \toolname{} construction and evaluation pipeline.}
\label{fig:pipeline}
\end{figure}

Figure~\ref{fig:pipeline} presents the \toolname{} workflow. Functions exercised by existing tests serve as injection contexts, providing concrete executions in which new bugs can be reached and observed. Each injection and transformation is validated by rebuilding the project and rerunning its construction-time witness. The pipeline comprises execution-context collection, buggy-code injection, and bug-preserving code transformation.

\suppressfloats[t]

\subsection{Execution Context Collection}

Given a real-world project and its existing test suite, \toolname{} first identifies the functions exercised by each test case.
This step provides a test-grounded search space for bug injection.
To collect the executed functions during runtime, \toolname{} applies program
instrumentation to the project.
Specifically, it parses the source code of the project and inserts logging calls at the
entry of each function.
During test execution, these logging calls record the executed functions and their source locations.
For a function $f$, we denote its source location as
$\ell(f) = (\mathrm{name}(f), \mathrm{path}(f), \mathrm{line}(f))$,
where $\mathrm{name}(f)$ is the function name, $\mathrm{path}(f)$ is the path of the
source file containing $f$, and $\mathrm{line}(f)$ is the starting line number of $f$.
Formally, executing a test case $t$ yields an ordered execution trace $\tau_t = [\ell_1, \ell_2, \ldots, \ell_n]$, where each $\ell_i$ is an observed function location. We define the corresponding execution context as the set of unique locations in that trace, $EC_t = \{\ell_i \mid \ell_i \in \tau_t\}$. Thus, $\tau_t$ preserves order, whereas $EC_t$ supports context membership and distinct-function counts.
Repeating this process for all test cases produces an \emph{execution context map},
$\mathcal{M}_{EC}$, which maps each test case to its execution context
$\mathcal{M}_{EC}(t) = EC_t$.

The execution context map serves as the basis for selecting candidate injection
locations.
\toolname{} restricts the injection search space to functions executed by a specific test case.
This design improves the likelihood that the injected bug is reachable and reduces the
cost of subsequent execution-guided injection.

\subsection{Buggy Code Injection}
\label{sec:bug_injection}

In the second stage, \toolname{} utilizes the collected execution contexts to synthesize
benchmark cases via buggy code injection.
We systematically cataloged test-observable bug patterns from CodeQL Java query examples~\citep{codeql_java_queries} and grouped them by the program property they violate: \emph{API Contract} for API contracts or specifications, \emph{Value Flow} for constraints on propagated values (e.g., dereferencing null values), and \emph{Logic} for domain-specific functional behavior. This grouping separates the contract, data-state, and functional reasoning needed to construct witnesses. To bound construction and agent-evaluation cost, we randomly selected ten eligible types while retaining examples from each category. The construction loop takes a bug-type specification, a test-reached execution context, and a replayable test supplying the input and oracle. Additional types use the same interface with a specification whose failure is observable; new ecosystems also require suitable tracing and replay adapters.
In our setting, accepted bugs are required to produce observable runtime failures when
exercised, which provides an automatic execution oracle for validating witnesses.
The bug types used in our benchmark construction are listed in Table~\ref{tab:bug_types} in Appendix~\ref{sec:implementation}.

Based on the above ingredients, the buggy code injection can be formulated as follows.
Given a test case $t$, an execution context $EC_t$, and a bug type $b$, \toolname{} leverages a coding agent to inject $b$ into the execution context $EC_t$ of test case $t$.
Since the functions in $EC_t$ are exercised, the injected bug is likely to be reachable by the construction-time witness derived from test case $t$.
After each injection attempt, \toolname{} rebuilds the modified project and reruns the
associated test case $t$.
If the test does not expose the bug, \toolname{} provides the execution feedback to the coding agent and requests a refinement. This process continues until the bug is observed or the attempt budget is exhausted.
This procedure retains cases whose injected bug is present in production code and empirically exposed by the construction-time witness.
The system prompt and user prompt used in buggy code injection are provided in Appendix~\ref{appendix:bug_injection_system_prompt} and Appendix~\ref{appendix:bug_injection_task_payload}.
~\looseness=-1

\subsection{Bug-Preserving Code Transformation}
\vspace{-5pt}
Since the bug types are derived from public CodeQL examples, directly reusing these snippets may produce benchmark cases that are syntactically or structurally similar
to publicly available bug types.
\toolname{} applies bug-preserving transformations that are
similar in spirit to code obfuscation.
Specifically, they alter the surface representation and
contextual structure of the injected bug while preserving its behavior under the same construction-time witness. The transformations reduce surface similarity to public bug examples and increase contextual complexity while preserving the injected bug.

Technically, \toolname{} is built on three families of transformation operators,
which manipulate call stack, data flow, and control flow structures.
The detailed definitions of the transformation operators are listed in Table~\ref{tab:operators} in Appendix~\ref{sec:implementation}.
Similar to buggy code injection,
the transformation stage is also execution-guided so that we can validate whether the transformed code preserves the injected bug.
If not, the transformation is reverted.
Through this process, \toolname{} can derive multiple benchmark variants from the same
injected bug.
These variants share an underlying bug type and construction-time witness while differing in call structure, data flow, and control flow. They provide repeated validation tasks under systematically varied structural contexts.
The prompt of the bug-preserving code transformation is given in Appendix~\ref{appendix:bug_preserving_transformation_task_payload}.

%% file: sections/evaluation.tex
\section{Evaluation}
\label{sec:evaluation}

We instantiate \toolname{} on six real-world Maven-based Java projects using ten observable bug types adapted from the CodeQL Java query library~\citep{codeql_java_queries}. The types span API Contract, Value Flow, and Logic bugs. The benchmark contains 1{,}300 buggy cases across four execution-context-length categories and four transformation-depth levels, and costs approximately \$16{,}500 to construct. Multiple cases per type provide observations across repositories and construction settings. Appendix~\ref{sec:implementation} and Appendix~\ref{app:benchmark_stats} give implementation details and benchmark statistics.

\subsection{Evaluation Setup}
\label{sec:evaluation_setup}

\paragraph{Bug-Validation Task.}
We evaluate every case under two settings that correspond to two important software-engineering scenarios. In the context-agnostic setting, denoted $\mathsf{NoEC}$, the agent receives the buggy repository, the natural language description of the bug type, the buggy location, but no execution context. This setting mimics the validation of static bug-detection results, where developers may have a localized bug report but no calling context showing how the buggy location is reached. In the context-guided setting, denoted $\mathsf{WithEC}$, the agent receives the same inputs together with the execution context. This setting mimics the bug-reproduction and debugging workflows, where a crash stack or recorded execution trace is available as a dynamic hint. 
The evaluation begins from a supplied bug report and location and measures the subsequent construction of an executable witness.
In the current benchmark instantiation, execution-context length ranges from 1 to 150 distinct production-side functions, grouped into four categories: \textit{short} ($[1,\,40]$), \textit{mid-short} ($[41,\,70]$), \textit{mid-long} ($[71,\,120]$), and \textit{long} ($[121,\,150]$).
Appendix~\ref{appendix:poc_generation_system_prompt} and Appendix~\ref{appendix:poc_generation_task_payload} provide the prompts used for executable bug validation.

\paragraph{Agent Settings.}
We evaluate six framework/model pairings: Codex and Claude Code with their hosted models, and OpenHands and OpenCode with four open-weight models. Table~\ref{tab:agents} lists the pairings and provider release identifiers~\citep{gpt54,claudesonnet45,deepseekv32,glm47,devstral2,qwen3coder480b}.

\Needspace{9\baselineskip}
\begingroup
\setlength{\intextsep}{3pt}
\setlength{\columnsep}{14pt}
\begin{wraptable}{r}{0.50\textwidth}
\vspace{-4pt}
\centering
\footnotesize
\setlength{\abovecaptionskip}{0pt}
\renewcommand{\arraystretch}{1.05}
\setlength{\tabcolsep}{2pt}
\caption{Evaluated framework/model pairings.}
\label{tab:agents}
\resizebox{\linewidth}{!}{%
\begin{tabular}{@{}cll@{}}
\toprule
\textbf{Agent} & \textbf{Framework} & \textbf{Model} \\
\midrule
A1 & Codex & GPT-5.4 \\
A2 & Claude Code & Claude Sonnet~4.5 \\
A3 & OpenHands & DeepSeek-V3.2 \\
A4 & OpenCode & ZAI GLM-4.7 \\
A5 & OpenHands & Devstral-2-123B \\
A6 & OpenCode & Qwen3-Coder-480B-A35B-Instruct \\
\bottomrule
\end{tabular}}
\end{wraptable}
Each case allows at most three validation attempts. Each attempt uses a 1{,}800-second agent timeout, a 1{,}200-second replay-verification timeout, a 600-second optional formatting timeout, and a 4{,}200-second case-level timeout. For OpenHands, we set the temperature to $0$ and the maximum output length to 4{,}096 tokens. Codex, Claude Code, and OpenCode use interface-default decoding because their evaluated interfaces do not expose directly comparable controls. Runs that exceed the case timeout, fail replay verification, or do not produce a repository test containing an executable witness are counted as failures. The evaluation costs approximately \$8{,}700.
\par\endgroup

\Needspace{15\baselineskip}
\subsection{RQ1: How Well Do Coding Agents Validate Bugs through Execution?}
\label{sec:overall}

\begingroup
\setlength{\intextsep}{3pt}
\setlength{\columnsep}{14pt}
\begin{wraptable}{r}{0.43\textwidth}
\vspace{-4pt}
\centering
\footnotesize
\setlength{\abovecaptionskip}{0pt}
\renewcommand{\arraystretch}{1.05}
\setlength{\tabcolsep}{3pt}
\caption{Agent validation success rates.}
\label{tab:overall}
\begin{tabular*}{\linewidth}{@{\extracolsep{\fill}}cccc@{}}
\toprule
\textbf{ID} & $\mathsf{NoEC}$ & $\mathsf{WithEC}$ & Average \\
\midrule
A1 & \ratecell{67.1} & \ratecell{70.7} & \ratecell{68.9} \\
A2 & \ratecell{57.1} & \ratecell{61.8} & \ratecell{59.4} \\
A3 & \ratecell{46.0} & \ratecell{49.8} & \ratecell{47.9} \\
A4 & \ratecell{44.2} & \ratecell{47.6} & \ratecell{45.9} \\
A5 & \ratecell{30.2} & \ratecell{33.8} & \ratecell{32.0} \\
A6 & \ratecell{17.1} & \ratecell{19.3} & \ratecell{18.2} \\
\bottomrule
\end{tabular*}
\end{wraptable}
To control evaluation cost, we randomly sample 900 cases from the 1{,}300-case benchmark and evaluate all six agents under both $\mathsf{NoEC}$ and $\mathsf{WithEC}$. Table~\ref{tab:overall} shows substantial performance differences. Codex (A1) and Claude Code (A2) achieve average validation success rates of 68.9\% and 59.4\%, while OpenHands with DeepSeek-V3.2 (A3) exceeds OpenCode with Qwen3-Coder-480B-A35B-Instruct (A6) by 29.7 percentage points. The best-performing agent still fails on 31.1\% of the cases.

Execution context improves validation success by 2.2--4.7 percentage points across the six agents.
These gains suggest that dynamic context helps agents identify relevant functions and paths for constructing executable witnesses.
However, the improvements do not change the agent ranking, and large performance gaps remain under $\mathsf{WithEC}$.
Even with execution context, the best-performing agent succeeds on 70.7\% of cases, leaving 29.3\% unsolved within the evaluation budget.
The supplied context can guide code inspection, but it does not provide a complete witness.
\par\endgroup

\mybox{
\textbf{Finding 1}: Current coding agents leave substantial room for improvement in executable bug validation, including when execution context is available.
}

\subsection{RQ2: How Do Bug Types Affect Executable Bug Validation?}
\label{sec:pattern}

To investigate the effect of bug types, we analyze a 500-case subset drawn from the full benchmark.
For each bug type, the subset contains 25 short-context cases and 25 long-context cases, all constructed with the same depth-3 transformation configuration and the same transformation operator sequence, and each case is evaluated under both the $\mathsf{NoEC}$ and $\mathsf{WithEC}$ settings.
As shown in Table~\ref{tab:per_pattern}, difficulty varies substantially across bug types and agents.
In particular, the Codex agent framework with GPT-5.4 (A1) and the Claude Code agent framework with Claude Sonnet~4.5 (A2) are relatively weak on Logic bugs, with $\mathsf{WithEC}$ success rates of 70.0\% and 60.0\%, respectively.
The open-source agents exhibit a different profile.
Their performance varies sharply across bug families, and they are generally less effective on Value Flow and Logic bugs than on API Contract bugs.

\begin{table}[!t]
\centering
\scriptsize
\renewcommand{\arraystretch}{1.12}
\setlength{\tabcolsep}{3.2pt}
\newcommand{\datacell}[1]{\cellcolor[HTML]{FFF3E6}#1}
\renewcommand{\weakcell}[1]{\cellcolor[HTML]{F6C77A}#1}
\caption{Successful bug validations across bug types. L and S denote long and short execution contexts. Each A/B cell reports successful cases under $\mathsf{WithEC}$ and $\mathsf{NoEC}$, respectively.}
\label{tab:per_pattern}
\resizebox{1.0\textwidth}{!}{%
\begin{tabular}{@{}lcccccccccccc@{}}
\toprule
\multirow{2}{*}{\raisebox{-0.6ex}{\textbf{Bug Type}}}
  & \multicolumn{2}{c}{\textbf{A1}}
  & \multicolumn{2}{c}{\textbf{A2}}
  & \multicolumn{2}{c}{\textbf{A3}}
  & \multicolumn{2}{c}{\textbf{A4}}
  & \multicolumn{2}{c}{\textbf{A5}}
  & \multicolumn{2}{c}{\textbf{A6}} \\
\cmidrule(lr){2-3}
\cmidrule(lr){4-5}
\cmidrule(lr){6-7}
\cmidrule(lr){8-9}
\cmidrule(lr){10-11}
\cmidrule(lr){12-13}
  & \textbf{L} & \textbf{S}
  & \textbf{L} & \textbf{S}
  & \textbf{L} & \textbf{S}
  & \textbf{L} & \textbf{S}
  & \textbf{L} & \textbf{S}
  & \textbf{L} & \textbf{S} \\
\midrule
Comparison of Identical Values
  & \datacell{18/17} & \datacell{22/21}
  & \datacell{17/14} & \datacell{19/16}
  & \datacell{14/10} & \datacell{15/15}
  & \datacell{11/11} & \datacell{15/13}
  & \datacell{9/7} & \datacell{10/9}
  & \datacell{4/4} & \datacell{6/5} \\
Typo in \texttt{equals}
  & \datacell{18/15} & \datacell{21/21}
  & \datacell{14/13} & \datacell{16/18}
  & \datacell{12/13} & \datacell{15/13}
  & \datacell{10/10} & \datacell{12/11}
  & \datacell{8/7} & \datacell{9/10}
  & \datacell{5/3} & \datacell{6/5} \\
Typo in \texttt{hashCode}
  & \datacell{16/13} & \datacell{18/17}
  & \datacell{16/11} & \datacell{19/14}
  & \datacell{10/9} & \datacell{15/12}
  & \datacell{11/8} & \datacell{13/14}
  & \datacell{7/6} & \datacell{10/7}
  & \datacell{4/4} & \datacell{5/5} \\
Hashed Value Without \texttt{hashCode}
  & \datacell{14/14} & \datacell{19/17}
  & \datacell{15/13} & \datacell{16/15}
  & \datacell{10/11} & \datacell{15/13}
  & \datacell{9/9} & \datacell{14/10}
  & \datacell{8/6} & \datacell{10/7}
  & \datacell{4/3} & \datacell{5/4} \\
Inconsistent \texttt{equals} and \texttt{hashCode}
  & \datacell{17/15} & \datacell{18/20}
  & \datacell{13/12} & \datacell{18/17}
  & \datacell{12/9} & \datacell{12/13}
  & \datacell{12/9} & \datacell{13/11}
  & \datacell{6/6} & \datacell{10/9}
  & \datacell{3/4} & \datacell{5/4} \\
\midrule
Intra-Function Null Dereference
  & \datacell{19/16} & \datacell{23/23}
  & \datacell{14/13} & \datacell{19/15}
  & \weakcell{9/9} & \weakcell{12/11}
  & \weakcell{11/9} & \weakcell{13/12}
  & \datacell{9/8} & \datacell{10/11}
  & \datacell{7/6} & \datacell{8/6} \\
Collection Element Null Dereference
  & \datacell{16/16} & \datacell{20/17}
  & \datacell{14/12} & \datacell{16/17}
  & \weakcell{10/8} & \weakcell{12/11}
  & \weakcell{8/7} & \weakcell{11/10}
  & \datacell{8/6} & \datacell{11/8}
  & \datacell{5/4} & \datacell{8/8} \\
\midrule
Guard Predicate Inversion
  & \weakcell{16/14} & \weakcell{18/16}
  & \weakcell{13/13} & \weakcell{17/14}
  & \datacell{13/10} & \datacell{17/13}
  & \datacell{12/9} & \datacell{14/13}
  & \weakcell{7/7} & \weakcell{8/8}
  & \weakcell{4/3} & \weakcell{5/5} \\
Skipped Aggregation Counter
  & \weakcell{16/18} & \weakcell{23/20}
  & \weakcell{15/12} & \weakcell{21/17}
  & \datacell{12/12} & \datacell{15/15}
  & \datacell{12/12} & \datacell{17/16}
  & \weakcell{8/5} & \weakcell{10/8}
  & \weakcell{4/4} & \weakcell{5/5} \\
In-Bounds Collection Misrouting
  & \weakcell{15/12} & \weakcell{17/18}
  & \weakcell{11/11} & \weakcell{13/15}
  & \datacell{12/12} & \datacell{14/15}
  & \datacell{12/11} & \datacell{14/13}
  & \weakcell{6/6} & \weakcell{8/7}
  & \weakcell{4/3} & \weakcell{5/5} \\
\bottomrule
\end{tabular}}
\end{table}

The gap between industrial and open-source agents varies by bug type.
In particular, OpenHands with DeepSeek-V3.2 (A3) generally outperforms the other evaluated open-source agents with open-weight models.
It narrows the gap to Claude Code with Claude Sonnet~4.5 (A2) on several Logic bug types, including Guard Predicate Inversion and In-Bounds Collection Misrouting.
Under $\mathsf{WithEC}$, A3 matches A2 on the former and slightly exceeds it on the latter.
However, this pattern does not extend to all bug types, as A3 still trails A2 on both Value Flow types.
These results show that open-source agent frameworks paired with strong open-weight models can approach industrial agents on specific bug types.

\Needspace{22\baselineskip}
\noindent\begin{minipage}[t]{0.60\textwidth}
\vspace{0pt}
We further quantify the effect of execution context across bug types.
Figure~\ref{fig:heatmap_diff} shows the difference in success rates between $\mathsf{WithEC}$ and $\mathsf{NoEC}$ for each agent--type pair.
We hold the depth-3 transformation configuration fixed across bug types.
The remaining differences therefore mainly reflect bug semantics and how well each agent uses the supplied context.
The gains from execution context vary across agents and bug types.
The remaining difficulty lies in reasoning about program properties within this context.
Agents must use the supplied information to choose concrete inputs, set up the required program state, and write checks that expose the faulty behavior.
Knowing which functions are involved does not by itself determine how to trigger the bug.
Different bug types require different forms of reasoning.
Agents and models therefore differ in how well they can turn the same contextual information into a working test.
Benchmarks need multiple bug categories and types that test program properties at different levels of detail and complexity.
\end{minipage}\hfill
\begin{minipage}[t]{0.36\textwidth}
  \vspace{0pt}
  \centering
  \includegraphics[width=\linewidth]{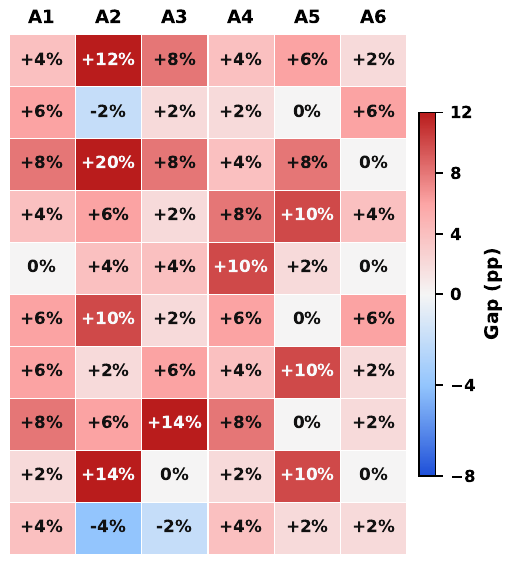}
  \setlength{\abovecaptionskip}{4pt}
  \captionof{figure}{Success-rate gain with execution context (percentage points). Types follow Table~\ref{tab:per_pattern}.}
  \label{fig:heatmap_diff}
\end{minipage}
\par

\mybox{
\textbf{Finding 2}: Validation performance varies by bug type. Logic bugs are generally challenging, while open-source agent frameworks with open-weight models approach industrial agents on several specific bug types.
}

\subsection{RQ3: How Does Bug-Validation Performance Vary with Execution-Context Length?}
\label{sec:exec_context}

To investigate the effect of execution context length, we fix the bug type to Guard Predicate Inversion and analyze a 400-case subset drawn from the full benchmark. The subset is stratified by four execution-context-length buckets and four transformation depth settings, with 25 cases in each length-by-depth cell. For each depth setting, all cases use the same transformation-operator sequence, and each case is evaluated under both the $\mathsf{NoEC}$ and $\mathsf{WithEC}$ settings.

\begin{figure}[t]
\centering
\includegraphics[width=.97\linewidth]{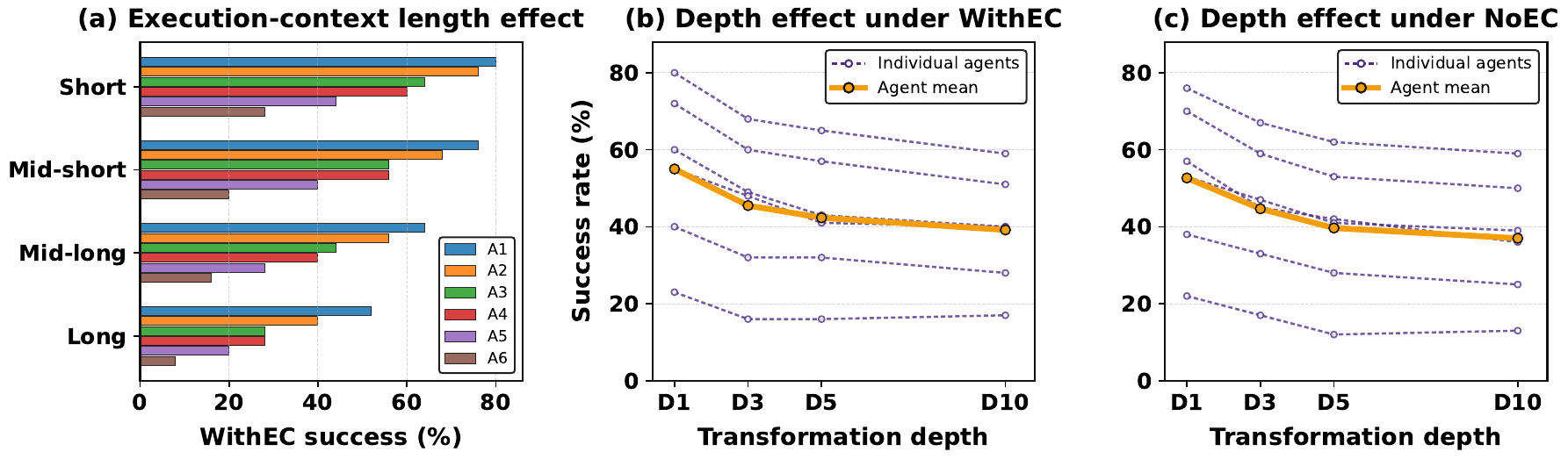}
\caption{The effect of execution context length (a) and transformation depth (b and c).}
\label{fig:context_transformation_combined}
\end{figure}

Figure~\ref{fig:context_transformation_combined}a summarizes the $\mathsf{WithEC}$ results, and Table~\ref{tab:exec_context} in Appendix~\ref{app:exec_context} reports both settings. Validation success is lower for longer execution contexts for every evaluated agent in both settings. Under $\mathsf{WithEC}$, Codex with GPT-5.4 (A1) decreases from 80\% on short contexts to 52\% on long contexts, Claude Code with Claude Sonnet~4.5 (A2) decreases by 36 percentage points, and OpenHands with Devstral-2-123B (A5) decreases from 44\% to 20\%. Execution context can narrow the relevant program region, while longer contexts still require agents to reconstruct more intermediate states, setup choices, and dependency chains before exposing the bug through a valid test.

\mybox{
\textbf{Finding 3}: Bug-validation success is substantially lower for longer execution contexts, including when execution context is provided.
}

\subsection{RQ4: How Does Bug-Validation Performance Vary with Transformation Depth?}
\label{sec:depth}

To investigate the effect of transformation depth, we fix the bug type to Guard Predicate Inversion and regroup the same 400-case subset by four transformation depths (1, 3, 5, and 10 applied transformations). Each depth-by-length cell contains 25 validated cases, and the transformation operator sequence is fixed. Each case is evaluated under both the $\mathsf{NoEC}$ and $\mathsf{WithEC}$ settings.

Figure~\ref{fig:context_transformation_combined}b and c show lower validation success at greater transformation depths in both settings. Under $\mathsf{WithEC}$, Codex with GPT-5.4 (A1) decreases from 80\% at depth~1 to 59\% at depth~10, Claude Code with Claude Sonnet~4.5 (A2) decreases from 72\% to 51\%, and OpenHands with DeepSeek-V3.2 (A3) decreases from 60\% to 40\%.

Deeper transformations can introduce additional code fragments, intermediate states, and dependency relations around the injected bug.
The agent must reason through the resulting implementation to construct a valid setup and oracle.
Success generally decreases with depth, with small plateaus or reversals in individual cells.
These comparisons characterize the evaluated transformation sequences rather than the independent contribution of each operator.
Table~\ref{tab:depth} in Appendix~\ref{app:depth} reports detailed statistics across transformation-depth buckets.
Appendix~\ref{sec:transform_appendix} provides further discussion of transformation composition and injection-stage patch evolution.

\mybox{
\textbf{Finding 4}: Greater transformation depth is generally associated with lower bug-validation success and provides a controllable benchmark-construction dimension.
}

\subsection{RQ5: How Realistic Are the Injected Bug Patches?}
\label{sec:realism}

The value of controlled bug injection depends on how closely its code changes resemble real development errors. We assess this resemblance through blinded source discrimination, using artifact-bearing positive controls to check judge sensitivity. We sample 700 of the 1{,}300 injected cases, stratified by repository and bug type, and match them to 700 historical bugs by repository and edit size. GPT-5.6-Terra and Claude Sonnet~4.6 independently see only blind IDs and production-code patches in the working-to-buggy direction.

Each judge performs \emph{single-patch judgment} three times per patch, predicting its source (4{,}200 judgments), and \emph{pairwise judgment} in both left--right orientations, selecting the more realistic patch (1{,}400 judgments). For 100 sampled pairs, we add conspicuous code artifacts to the injected patch while preserving its bug behavior and repeat the pairwise task (200 judgments). All calls use independent contexts, giving 11{,}600 judgments across both models.

Pairwise accuracy is the primary equivalence endpoint; single-patch accuracy and area under the receiver operating characteristic curve (ROC AUC) are auxiliary endpoints. Chance corresponds to 0.5. We use 10{,}000 whole-cluster bootstrap resamples, retaining related judgments together, and require each endpoint's complete 90\% confidence interval (CI) to lie within 0.45--0.55~\citep{lakens2017equivalence,fieldwelsh2007bootstrap}. Positive controls require accuracy of at least 80\% and a 95\% CI above 50\%. Appendix~\ref{app:realism} details the protocol and diagnostics.

\begin{figure}[!htbp]
  \centering
  \includegraphics[width=.9\linewidth]{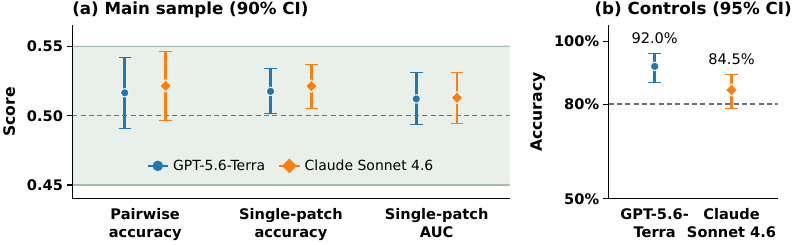}
  \caption{Blinded patch judgments. (a) Main-sample estimates and 90\% cluster CIs; shading marks the 0.45--0.55 equivalence range and the dashed line marks chance. (b) Artifact-bearing controls with 95\% CIs and the 80\% accuracy threshold.}
  \label{fig:realism_results}
\end{figure}

Figure~\ref{fig:realism_results} reports pairwise accuracies of 51.64\% and 52.14\%. Both judges meet the primary and auxiliary equivalence criteria. Their positive-control accuracies of 92.00\% and 84.50\% show sensitivity to conspicuous artifacts. The supplementary pairwise AUC interval for Claude Sonnet~4.6 slightly exceeds the equivalence bound; Appendix~\ref{app:realism} reports this result and position preferences.

\mybox{\textbf{Finding 5}: Injected patches resemble historical bugs in blinded comparisons, combining code-level realism with controlled construction for evaluating executable bug validation.}

%% file: sections/related.tex
\section{Related Work}
\vspace{-5pt}

\noindent\emph{\textbf{Input Generation.}}
Randoop, EvoSuite, and DART generate tests using execution feedback, coverage objectives, or path constraints~\citep{randoop2007,evosuite2011,dart2005}. KATCH, AFLGo, WAFLGo, CAFL, and SDFuzz direct execution using locations, patches, data constraints, or vulnerability states~\citep{katch2013,aflgo2017,waflgo2024,cafl2021,sdfuzz2024}. Target-directed input generation shares the need to satisfy the conditions that expose a bug. \toolname{} evaluates repository cases with supplied bug descriptions and locations, asking agents to construct both the input and its execution harness.

\noindent\emph{\textbf{Benchmarks.}}
LIBRO, Issue2Test, and BRT Agent generate reproducing tests from bug reports~\citep{libro2023,issue2test2026,cheng2025agentic}; AnyPoC validates candidate reports, and Differential Prompting uses program pairs to generate failure-inducing tests~\citep{zhao2026anypoc,differentialprompting2023}. CyberGym~\citep{wang2026cybergym}, SEC-bench~\citep{lee2025secbench}, SEC-bench Pro~\citep{secbenchpro2026}, and exploit generators~\citep{DBLP:journals/corr/abs-2506-04962,poco2025,v2e2026} evaluate vulnerability reproduction; SecCodeBench-V2 evaluates secure code generation and repair~\citep{seccodebenchv2}. Defects4J curates fault--test pairs, while LAVA, EvilCoder, and FixReverter inject bugs~\citep{just2014defects4j,dolangavitt2016lava,pewny2016evilcoder,fixreverter2022}. \toolname{} creates multi-category targets with construction-time witnesses and controlled structural variants.

\noindent\emph{\textbf{Task and Data Construction.}}
Change2Task derives verified coding-agent tasks from repository history across five maintenance task families~\citep{qi2026change2task}. Related work on data governance studies how heterogeneous sources can be selected and combined during multilingual model adaptation~\citep{qi2025governance}. \toolname{} complements these efforts by constructing fresh, execution-validated targets specifically for bug-witness construction.

%% file: sections/conclusion.tex
\section{Conclusion}
\vspace{-5pt}
This paper presents \toolname{}, an automated framework for constructing benchmarks for executable bug validation. It injects bugs into real projects and retains cases with a witness that exposes the faulty behavior. Bug-preserving transformations vary the surrounding program structure, allowing validation to be studied across different contextual demands. Our evaluation shows that coding agents struggle to construct witnesses as execution contexts grow and transformations deepen. Blinded comparisons also find the injected patches difficult to distinguish from historical bugs under the evaluated protocol. \toolname{} offers a way to generate fresh validation tasks as agents evolve and study progress toward more reliable AI-assisted code auditing.

%% file: sections/appendix.tex
\appendix

\input{appendix/limitations}

\input{appendix/impact}

\input{appendix/implementation}

\input{appendix/statistics}

\input{appendix/context_depth}

\input{appendix/operator}

\FloatBarrier
\input{appendix/realism}

\input{appendix/prompt}

%% file: appendix/limitations.tex
\section{Limitations}
\label{app:limitations}
The current implementation covers ten observable bug types in Maven-based Java projects with replayable tests. Additional bug specifications and language-specific tracing and replay adapters would extend this coverage. Context-length and transformation-depth comparisons use a fixed bug type to control the target semantics; broader combinations would test how these factors interact across types. The blinded patch study characterizes source discrimination under the two reported judges and presentation protocol. Broader model coverage and developer assessments could examine other aspects of realism, including how injected bugs resemble those encountered during software maintenance.

%% file: appendix/impact.tex
\section{Broader Impact}
\label{sec:broader_impact}

This work advances the evaluation of AI coding agents on executable bug validation in code-auditing scenarios. The injected bugs are isolated benchmark artifacts and are not changes proposed for deployed software. The framework can support research on evidence-grounded code auditing and on coding agents that validate reported bugs through execution.

Bug-validation performance varies with execution context length, transformation depth, and bug type across the evaluated agents. These dimensions support construction of cases with different reasoning requirements and difficulty levels.

The current implementation focuses on Maven-based Java projects and bug types derived from CodeQL Java queries. Applying the workflow to another ecosystem requires a runnable test environment, observable validation signals, and language-specific adapters. The benchmark methodology can support further study of bug validation for AI-assisted software engineering and security analysis.

%% file: appendix/implementation.tex
\section{Implementation Details}
\label{sec:implementation}

\paragraph{Project and Bug Type Preparation.}
We instantiate \toolname{} on six real-world Java projects spanning distributed middleware, data serialization, and multipurpose utility libraries. These projects provide injection sites across varied execution contexts.
All selected projects use Maven as their build system, which allows an individual test
case to be replayed with a localized command such as:
\[
\texttt{./mvnw -q -pl <module> -Dtest=<test-class> test}.
\]
We exclude tests that depend on external services, network access, nondeterministic
timing, or excessive runtime.
Each experiment starts from a clean repository template, and all injection,
transformation, and evaluation runs are performed in isolated working copies.
The current implementation targets Maven-based Java projects. Adapting the workflow to another language or build system requires isolated execution of a designated test and corresponding tracing, transformation, and validation adapters.

For bug type selection, we require the injected bug to produce an observable failure, such as an assertion failure or a runtime exception.
Types whose effects are difficult to observe deterministically, such as pure performance regressions, are outside the scope of this benchmark.
Among the eligible patterns surveyed from CodeQL Java query examples, we randomly selected ten types across the three categories to bound construction and evaluation cost.
Representative types include Guard Predicate Inversion, Intra-Function Null
Dereference, and Comparison of Identical Values.
Table~\ref{tab:bug_types} lists all ten types organized by category.

\begin{table}[t]
\centering
\small
\renewcommand{\arraystretch}{1.12}
\setlength{\tabcolsep}{5pt}
\caption{The ten bug types used in the benchmark construction.}
\label{tab:bug_types}
\begin{tabular}{>{\centering\arraybackslash}m{1.8cm}lp{6.0cm}}
\toprule
\textbf{Category} & \textbf{Type} & \textbf{Bug-type description} \\
\midrule
\multirow[c]{15}{*}{\shortstack[c]{API\\Contract}}
  & Comparison of Identical Values
  & A condition compares a value with itself, producing a constant result that silently disables validation or equality logic. \\
  & Typo in \texttt{equals}
  & A function overriding \texttt{equals(Object)} has the wrong name or signature, so Java keeps using reference equality. \\
  & Typo in \texttt{hashCode}
  & A function intended to override \texttt{hashCode()} has the wrong name or signature, so hashed collections use identity hashing. \\
  & Hashed Value Without \texttt{hashCode}
  & A class overrides \texttt{equals(Object)} but not \texttt{hashCode()}, so logically equal keys may hash to different buckets. \\
  & Inconsistent \texttt{equals} and \texttt{hashCode}
  & \texttt{equals} and \texttt{hashCode} use different fields or unstable state, breaking lookup in \texttt{HashMap} or \texttt{HashSet}. \\
\midrule
\multirow[c]{6}{*}{\shortstack[c]{Value\\Flow}}
  & Intra-Function Null Dereference
  & A value that was checked for null is later overwritten, moved outside the guard, or dereferenced on an unguarded path. \\
  & Collection Element Null Dereference
  & Maintenance or eviction leaves a null element in a list, map, or cache, and a later lookup or iteration assumes it is non-null. \\
\midrule
\multirow[c]{9}{*}{Logic}
  & Guard Predicate Inversion
  & A guard condition is flipped or weakened, sending valid inputs to an error or skip path, or letting invalid inputs continue. \\
  & Skipped Aggregation Counter
  & A branch still processes an item but skips the counter or total update, so returned metrics undercount the work performed. \\
  & In-Bounds Collection Misrouting
  & A loop writes an element to a legal but wrong index or key, leaving the collection complete but logically misordered. \\
\bottomrule
\end{tabular}
\end{table}

\begin{table}[t]
\centering
\small
\renewcommand{\arraystretch}{1.12}
\setlength{\tabcolsep}{5pt}
\caption{The ten bug-preserving transformation operators.}
\label{tab:operators}
\begin{tabular}{llp{6.0cm}}
\toprule
\textbf{Family} & \textbf{Operator} & \textbf{Effect} \\
\midrule
\multirow{8}{*}{Call Stack}
  & Helper Call Insertion
  & Adds an internal helper or wrapper call between the original caller and the bug site, increasing call-chain depth. \\
  & Call-Site Rerouting
  & Replaces a direct internal call with a forwarding function, so the same logic is reached through an extra dispatch step. \\
  & Extract-and-Apply Split
  & Splits one computation into an extraction helper and an application helper, forcing the bug to cross function boundaries. \\
\midrule
\multirow{15}{*}{Data Flow}
  & Parameter Bundling
  & Packs related arguments or locals into an internal object that is passed through helper calls instead of using separate values. \\
  & Local State Bundling
  & Moves related local variables into a context object, so later code reads state through fields rather than direct locals. \\
  & Intermediate State Passing
  & Stores intermediate values in a state holder and reads them later across helper calls, lengthening the value flow path. \\
  & Cached Value Access
  & Routes reads of a derived value through a cached accessor instead of direct recomputation or direct field access. \\
  & Nested Data Access
  & Replaces direct value access with access through a nested container, wrapper, map entry, or list element. \\
  & Pipeline Stage Split
  & Breaks a multi-step function into ordered internal stages, making the bug appear inside a staged processing pipeline. \\
\midrule
\multirow{2}{*}{Control Flow}
  & Rare Input Path Conditioning
  & Restricts the bug path to rare but deterministic input conditions, preserving normal behavior for common inputs. \\
\bottomrule
\end{tabular}
\end{table}

\paragraph{Execution Context Collection.}
For each selected test case, we collect the ordered production-side execution path
exercised by the test and summarize it as its execution context.
To study how execution-context length affects bug validation, we stratify the collected execution contexts into length-based buckets.
Here, execution context length refers to the number of distinct production-side functions
reached during the test run, as recorded in the execution log
$\tau_t$. Specifically, we use four buckets: \textit{short}, \textit{mid-short},
\textit{mid-long}, and \textit{long}, with execution context lengths of $[1,\,40]$,
$[41,\,70]$, $[71,\,120]$, and $[121,\,150]$ production functions,
respectively. Shorter contexts correspond to more constrained execution paths.
Longer contexts capture broader execution paths that may traverse multiple production
functions, classes, or modules.

\paragraph{Buggy Code Injection.}
We perform bug injection with Claude Code backed by Claude Sonnet~4.5.
For each candidate case, the injection process is allowed up to three attempts.
Before every attempt, the repository is restored to a clean workspace so that failed
attempts do not accumulate state.
A candidate is archived only if it satisfies three sequential validation checks.
First, it must introduce a genuine production-code diff; cases with no change,
formatting-only edits, or modifications confined to test or build files are rejected.
Second, the diff must be structurally consistent with the intended bug type; we use
lightweight syntactic checks to filter out obvious mismatches, such as injecting a
null-pointer bug when the requested type is an equality inconsistency.
Third, the replay command must pass on the clean baseline and expose the injected bug
on the modified repository; cases that fail to compile, fail for unrelated infrastructure reasons, or do not expose the intended bug are discarded.

\paragraph{Bug-Preserving Transformation.}
After obtaining a valid injected case, we apply bug-preserving transformations, also using
Claude Code backed by Claude Sonnet~4.5. We allow one attempt per transformation.
To diversify the structural embedding of injected bugs, we define ten transformation
operators across three families.
Table~\ref{tab:operators} lists all ten operators.
Each transformation prompt instructs the agent to rewrite the region containing the injected bug while preserving the construction-time witness. We apply 1, 3, 5, or 10 operators sequentially after injection. Each step rebuilds the project and reruns the replay command. A transformed case is retained only when the project remains valid and the same witness still exposes the injected bug. The transformations therefore vary program structure without repairing the bug.

%% file: appendix/statistics.tex
\section{Detailed Benchmark Statistics}
\label{app:benchmark_stats}

\begin{figure}[t]
\centering
\includegraphics[width=1.0\textwidth]{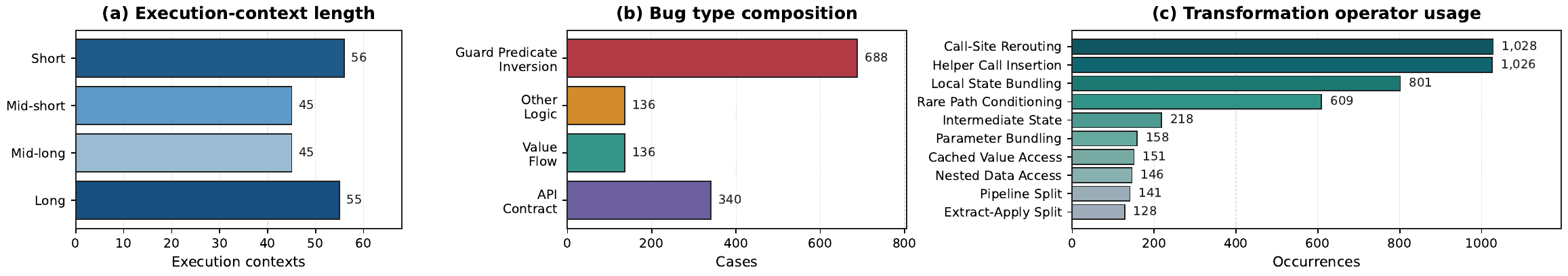}
\caption{The distributions of execution context length, different categories of bug types, and the occurrence of transformation operators.}
\label{fig:benchmark_stats}
\end{figure}

The constructed benchmark contains 1{,}300 archived buggy cases from six real-world
Java projects. Each case stores a buggy repository snapshot, the execution context associated with the selected test, the applied bug type, the transformation sequence, and the construction-time witness used for validation. Figure~\ref{fig:benchmark_stats} summarizes the benchmark composition. At the case level, the benchmark contains 201 execution contexts: 56 short, 45 mid-short, 45 mid-long, and 55 long contexts.

The benchmark covers all ten bug types in Table~\ref{tab:bug_types}.
Guard Predicate Inversion appears most frequently, with 688 cases, because it is used
as the anchor type for systematically varying transformation depth and execution context length. The remaining nine types each appear in 68 cases, yielding balanced coverage over
the other Logic, Value Flow, and API Contract bugs. At the category level, the benchmark therefore contains 824 Logic cases, 136 Value Flow cases, and 340 API Contract cases.
Across all cases, the ten transformation operators in Table~\ref{tab:operators}
appear 4{,}406 times in total. The most frequent operators are Call-Site Rerouting (1{,}028 occurrences), Helper Call Insertion (1{,}026), Local State Bundling (801), and Rare Input Path Conditioning (609), followed by Intermediate State Passing (218), Parameter Bundling (158), Cached Value Access (151), Nested Data Access (146), Pipeline Stage Split (141), and
Extract-and-Apply Split (128).

Approximately 63\% of attempted candidates pass the construction pipeline. The estimated cost includes failed attempts and successful bug injection, transformation, execution-context collection, replay validation, and construction-time witness validation. The end-to-end construction cost is approximately \$16{,}500.

%% file: appendix/context_depth.tex
\section{Extended Analysis of Execution Context Effects}
\label{app:exec_context}

\newcommand{\ppbasemaxdelta}{36}
\newcommand{\setppbasemax}[1]{\gdef\ppbasemaxdelta{#1}}
\newcommand{\ppbasedownmark}[1]{\raisebox{-0.55ex}{\scriptsize\textcolor{red!72!black}{$\downarrow$#1}}}
\newcommand{\ppbaseupmark}[1]{\raisebox{-0.55ex}{\scriptsize\textcolor{blue!55!black}{$\uparrow$#1}}}
\newcommand{\ppbaseflatmark}{\raisebox{-0.55ex}{\scriptsize\textcolor{black!55}{0pp}}}
\newcommand{\ppbasedowncell}[2]{%
  \begingroup
  \pgfmathsetmacro{\s}{min(abs(#1)/\ppbasemaxdelta,1)}%
  \pgfmathsetmacro{\r}{1.00}%
  \pgfmathsetmacro{\g}{1.00 - 0.30*\s}%
  \pgfmathsetmacro{\b}{1.00 - 0.36*\s}%
  \edef\temp{\noexpand\cellcolor[rgb]{\r,\g,\b}\noexpand\strut\unexpanded{#2\,\ppbasedownmark{#1pp}}}%
  \temp
  \endgroup
}
\newcommand{\ppbaseupcell}[2]{%
  \cellcolor[rgb]{0.92,0.96,1.00}\strut #2\,\ppbaseupmark{#1pp}%
}
\newcommand{\ppbaseflatcell}[1]{%
  \cellcolor[rgb]{1,1,1}\strut #1\,\ppbaseflatmark%
}

\begin{table}[t]
\centering
\footnotesize
\renewcommand{\arraystretch}{1.10}
\setlength{\tabcolsep}{0pt}
\setppbasemax{36}
\caption{Bug-validation success rates across execution context length buckets. Subscript annotations give percentage-point differences from the short bucket.}
\label{tab:exec_context}
\begin{tabular*}{1.0\textwidth}{@{}
M{0.10\textwidth}
M{0.1125\textwidth}M{0.1125\textwidth}
M{0.1125\textwidth}M{0.1125\textwidth}
M{0.1125\textwidth}M{0.1125\textwidth}
M{0.1125\textwidth}M{0.1125\textwidth}
@{}}
\toprule
\multirow{2}{*}{\raisebox{-0.6ex}{\textbf{Agent ID}}}
  & \multicolumn{2}{c}{\textbf{Short}}
  & \multicolumn{2}{c}{\textbf{Mid-Short}}
  & \multicolumn{2}{c}{\textbf{Mid-Long}}
  & \multicolumn{2}{c}{\textbf{Long}} \\
\cmidrule(lr){2-3}
\cmidrule(lr){4-5}
\cmidrule(lr){6-7}
\cmidrule(lr){8-9}
  & \textbf{NoEC} & \textbf{WithEC}
  & \textbf{NoEC} & \textbf{WithEC}
  & \textbf{NoEC} & \textbf{WithEC}
  & \textbf{NoEC} & \textbf{WithEC} \\
\midrule
A1
  & 80\% & 80\%
  & \ppbasedowncell{8}{72\%} & \ppbasedowncell{4}{76\%}
  & \ppbasedowncell{16}{64\%} & \ppbasedowncell{16}{64\%}
  & \ppbasedowncell{32}{48\%} & \ppbasedowncell{28}{52\%} \\
\midrule[0.25pt]
A2
  & 72\% & 76\%
  & \ppbasedowncell{8}{64\%} & \ppbasedowncell{8}{68\%}
  & \ppbasedowncell{16}{56\%} & \ppbasedowncell{20}{56\%}
  & \ppbasedowncell{32}{40\%} & \ppbasedowncell{36}{40\%} \\
\midrule[0.25pt]
A3
  & 60\% & 64\%
  & \ppbasedowncell{8}{52\%} & \ppbasedowncell{8}{56\%}
  & \ppbasedowncell{20}{40\%} & \ppbasedowncell{20}{44\%}
  & \ppbasedowncell{32}{28\%} & \ppbasedowncell{36}{28\%} \\
\midrule[0.25pt]
A4
  & 60\% & 60\%
  & \ppbasedowncell{8}{52\%} & \ppbasedowncell{4}{56\%}
  & \ppbasedowncell{20}{40\%} & \ppbasedowncell{20}{40\%}
  & \ppbasedowncell{32}{28\%} & \ppbasedowncell{32}{28\%} \\
\midrule[0.25pt]
A5
  & 44\% & 44\%
  & \ppbasedowncell{8}{36\%} & \ppbasedowncell{4}{40\%}
  & \ppbasedowncell{16}{28\%} & \ppbasedowncell{16}{28\%}
  & \ppbasedowncell{28}{16\%} & \ppbasedowncell{24}{20\%} \\
\midrule[0.25pt]
A6
  & 24\% & 28\%
  & \ppbasedowncell{4}{20\%} & \ppbasedowncell{8}{20\%}
  & \ppbasedowncell{12}{12\%} & \ppbasedowncell{12}{16\%}
  & \ppbasedowncell{16}{8\%} & \ppbasedowncell{20}{8\%} \\
\bottomrule
\end{tabular*}
\end{table}

This appendix expands RQ3 using the same 400-case controlled subset of Guard Predicate Inversion used in the main text. The subset contains four execution context length buckets and four transformation depth buckets, with 25 validated cases for each bucket combination. Table~\ref{tab:exec_context} groups these cases by execution context length and aggregates over transformation depths.

Table~\ref{tab:exec_context} gives a finer-grained view of how execution-context length affects validation. Success decreases with longer contexts for every agent under both $\mathsf{NoEC}$ and $\mathsf{WithEC}$. Under $\mathsf{WithEC}$, A1 drops from 80\% to 52\%, A2 from 76\% to 40\%, A3 from 64\% to 28\%, and A4 from 60\% to 28\% between the short and long buckets.

Execution-context guidance improves several buckets, while success still declines as contexts grow. Under $\mathsf{WithEC}$, A1 loses 28 percentage points and A2 loses 36 percentage points from short to long contexts. Longer contexts require agents to select relevant setup operations, preserve intermediate states, and construct an oracle that exposes the bug.

The decline is often progressive across buckets. Longer execution contexts expand the program behavior that a witness must reconstruct before the bug becomes observable.

\section{Extended Analysis of Transformation Depth Effects}
\label{app:depth}

This appendix expands RQ4 using the same 400-case controlled subset. The subset contains four transformation depth buckets and four execution context length buckets, with 25 validated cases for each bucket combination. Table~\ref{tab:depth} groups these cases by transformation depth and aggregates over execution context lengths.

Table~\ref{tab:depth} expands the transformation-depth analysis. Under $\mathsf{WithEC}$, success from depth~1 to depth~10 decreases from 80\% to 59\% for A1, 72\% to 51\% for A2, 60\% to 40\% for A3, and 55\% to 40\% for A4. Transformation depth therefore provides a practical dimension for adjusting bug-validation difficulty.

\begin{table}[t]
\centering
\footnotesize
\renewcommand{\arraystretch}{1.10}
\setlength{\tabcolsep}{0pt}
\setppbasemax{21}
\caption{Bug-validation success rates across transformation-depth buckets. Subscript annotations give percentage-point differences from depth~1.}
\label{tab:depth}
\begin{tabular*}{1.0\textwidth}{@{}
M{0.10\textwidth}
M{0.1125\textwidth}M{0.1125\textwidth}
M{0.1125\textwidth}M{0.1125\textwidth}
M{0.1125\textwidth}M{0.1125\textwidth}
M{0.1125\textwidth}M{0.1125\textwidth}
@{}}
\toprule
\multirow{2}{*}{\raisebox{-0.6ex}{\textbf{Agent ID}}}
  & \multicolumn{2}{c}{\textbf{Depth 1}}
  & \multicolumn{2}{c}{\textbf{Depth 3}}
  & \multicolumn{2}{c}{\textbf{Depth 5}}
  & \multicolumn{2}{c}{\textbf{Depth 10}} \\
\cmidrule(lr){2-3}
\cmidrule(lr){4-5}
\cmidrule(lr){6-7}
\cmidrule(lr){8-9}
  & \textbf{NoEC} & \textbf{WithEC}
  & \textbf{NoEC} & \textbf{WithEC}
  & \textbf{NoEC} & \textbf{WithEC}
  & \textbf{NoEC} & \textbf{WithEC} \\
\midrule
A1
  & 76\% & 80\%
  & \ppbasedowncell{9}{67\%} & \ppbasedowncell{12}{68\%}
  & \ppbasedowncell{14}{62\%} & \ppbasedowncell{15}{65\%}
  & \ppbasedowncell{17}{59\%} & \ppbasedowncell{21}{59\%} \\
\midrule[0.25pt]
A2
  & 70\% & 72\%
  & \ppbasedowncell{11}{59\%} & \ppbasedowncell{12}{60\%}
  & \ppbasedowncell{17}{53\%} & \ppbasedowncell{15}{57\%}
  & \ppbasedowncell{20}{50\%} & \ppbasedowncell{21}{51\%} \\
\midrule[0.25pt]
A3
  & 57\% & 60\%
  & \ppbasedowncell{12}{45\%} & \ppbasedowncell{11}{49\%}
  & \ppbasedowncell{15}{42\%} & \ppbasedowncell{17}{43\%}
  & \ppbasedowncell{21}{36\%} & \ppbasedowncell{20}{40\%} \\
\midrule[0.25pt]
A4
  & 53\% & 55\%
  & \ppbasedowncell{6}{47\%} & \ppbasedowncell{7}{48\%}
  & \ppbasedowncell{12}{41\%} & \ppbasedowncell{14}{41\%}
  & \ppbasedowncell{14}{39\%} & \ppbasedowncell{15}{40\%} \\
\midrule[0.25pt]
A5
  & 38\% & 40\%
  & \ppbasedowncell{5}{33\%} & \ppbasedowncell{8}{32\%}
  & \ppbasedowncell{10}{28\%} & \ppbasedowncell{8}{32\%}
  & \ppbasedowncell{13}{25\%} & \ppbasedowncell{12}{28\%} \\
\midrule[0.25pt]
A6
  & 22\% & 23\%
  & \ppbasedowncell{5}{17\%} & \ppbasedowncell{7}{16\%}
  & \ppbasedowncell{10}{12\%} & \ppbasedowncell{7}{16\%}
  & \ppbasedowncell{9}{13\%} & \ppbasedowncell{6}{17\%} \\
\bottomrule
\end{tabular*}
\end{table}

The effect of transformation depth differs from the effect of execution context length. Execution context length increases the amount of triggering behavior that the test must reconstruct, whereas transformation depth changes how the injected bug is structurally embedded in the code. As more transformations are applied, the same underlying bug may be surrounded by additional code fragments, modified control structure, or altered data dependencies. Consequently, the agent must reason through a transformed implementation before it can synthesize a valid setup and oracle. This makes transformation depth a construction-side complexity factor that complements execution context length.

Transformation depth yields an overall difficulty trend with small plateaus and reversals in individual cells. For example, under $\mathsf{WithEC}$, OpenHands with Devstral-2-123B (A5) remains at 32\% from depth~3 to depth~5, and OpenCode with Qwen3-Coder-480B-A35B-Instruct (A6) increases from 16\% at depth~5 to 17\% at depth~10. Transformations are compositional: each additional step can affect code shape, control flow, and data flow in different ways. Transformation count thus provides a coarse-grained difficulty control; finer-grained metrics would account for the structural and semantic effects of individual operators.

%% file: appendix/operator.tex
\section{Additional Analysis of Transformation Dimension Effects}
\label{sec:transform_appendix}

\paragraph{Transformation composition.}
The transformation families change different aspects of the code surrounding an injected bug.
Call-stack operators introduce helper boundaries or additional dispatch steps.
Data-flow operators change how values are stored, accessed, and passed between functions, while control-flow operators change the conditions under which the buggy path is reached.
These changes can affect the program state and call sequence that an agent must reconstruct before a test exposes the bug.

A transformed case may contain multiple operators, and later steps can reshape code introduced by earlier ones.
The depth comparisons in RQ4 therefore concern the evaluated transformation sequences rather than the independent contribution of each operator.
Grouping cases by operator occurrence alone does not separate these contributions.
We examine the resulting patches below to describe how their size, structure, and retained content vary with transformation depth.

\paragraph{Injection-stage patch evolution.}
We analyze the injection-stage artifacts to examine how transformation depth changes the structure of the generated buggy patches.
This analysis uses the same 400-case transformation-depth subset as RQ4, with one fixed bug type, four transformation-depth buckets, and 100 successfully validated injected cases per bucket.
Across these cases, we analyze 1{,}900 step-level transformation patches and the final buggy patch retained for each case.
The analysis complements the validation results by describing the code changes produced during construction.

\begin{figure}[t]
\centering
\includegraphics[width=1.0\textwidth]{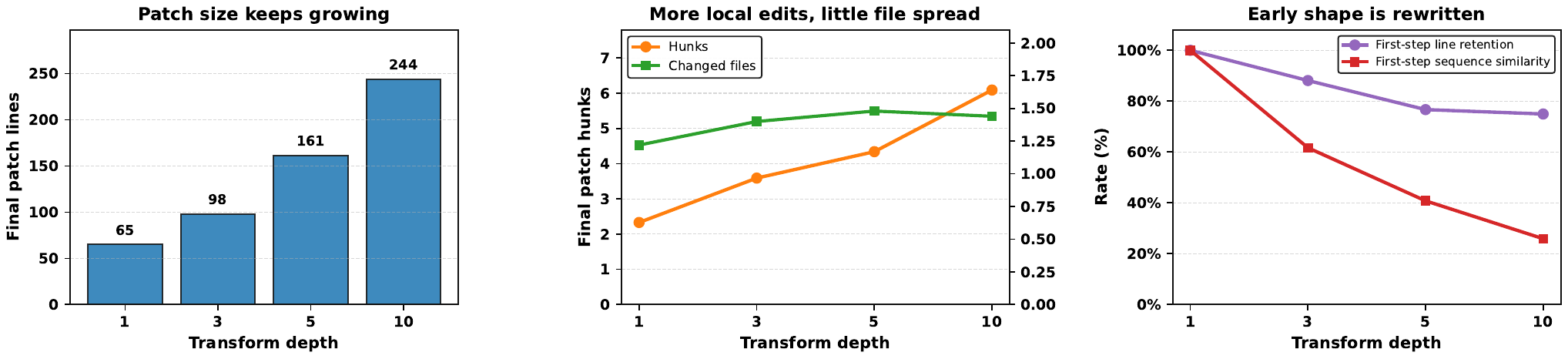}
\caption{Injection-stage patch evolution by transformation depth. Final patch size and hunk count grow with depth, while the number of changed files remains nearly flat.}
\label{fig:inject_depth_complexity}
\end{figure}

\begin{figure}[t]
\centering
\includegraphics[width=1.0\textwidth]{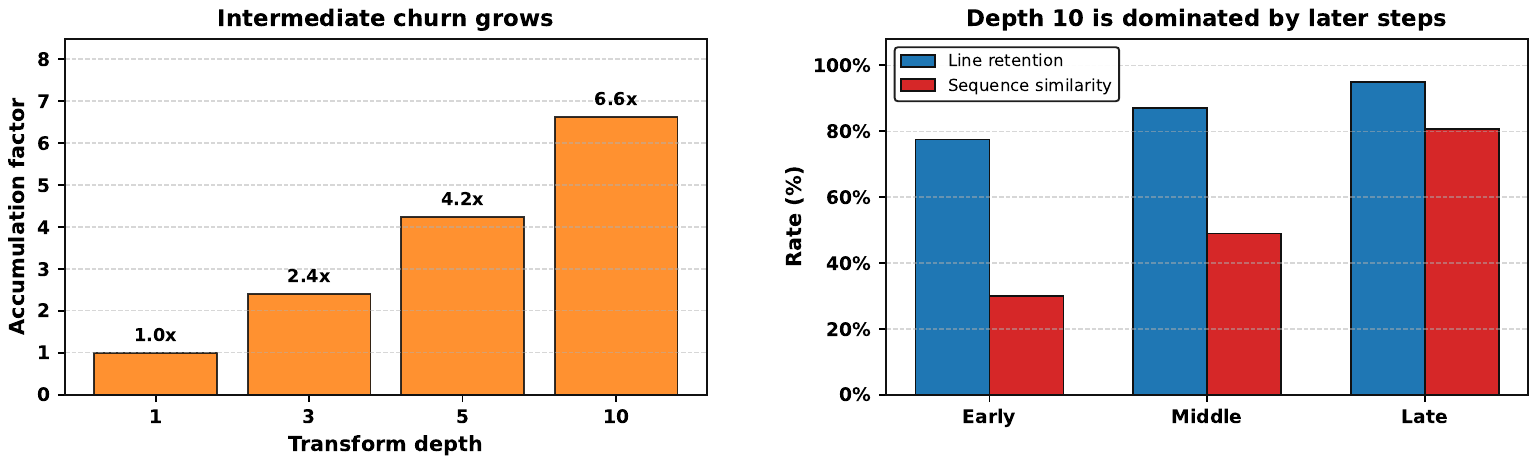}
\caption{Intermediate edit churn and step-position effects. At depth~10, later transformation steps have higher line retention and sequence similarity to the final buggy patch than earlier steps.}
\label{fig:inject_depth_retention}
\end{figure}

\begin{table}[t]
\centering
\footnotesize
\renewcommand{\arraystretch}{1.08}
\setlength{\tabcolsep}{5pt}
\caption{Injection-stage patch-evolution metrics by transformation depth. Line accumulation is the ratio between cumulative step-patch lines and final-patch lines. Retention and sequence similarity are measured from the first transformation step to the final buggy patch.}
\label{tab:inject_depth_complexity}
\resizebox{1.0\textwidth}{!}{
\begin{tabular}{@{}crrrrrrr@{}}
\toprule
\textbf{Depth} & \textbf{Runs} & \textbf{Final lines} & \textbf{Hunks} & \textbf{Changed files} & \textbf{Line accum.} & \textbf{First-step retention} & \textbf{First-step similarity} \\
\midrule
1  & 100 & 65.0  & 2.33 & 1.22 & 1.00$\times$ & 100.0\% & 100.0\% \\
3  & 100 & 97.8  & 3.59 & 1.40 & 2.40$\times$ & 88.1\%  & 61.7\% \\
5  & 100 & 161.5 & 4.34 & 1.48 & 4.25$\times$ & 76.7\%  & 40.8\% \\
10 & 100 & 243.7 & 6.09 & 1.44 & 6.63$\times$ & 74.9\%  & 25.9\% \\
\bottomrule
\end{tabular}}
\end{table}

Figure~\ref{fig:inject_depth_complexity} and Table~\ref{tab:inject_depth_complexity} show that greater transformation depth is associated with larger final patches and more hunks, without substantially spreading edits across more files.
The mean final patch size grows from 65.0 non-empty diff lines at depth~1 to 243.7 at depth~10, while the mean hunk count grows from 2.33 to 6.09.
In contrast, the mean number of changed files remains nearly flat, from 1.22 to 1.44.
The additional edits therefore remain concentrated within a small number of files, even as the final patch becomes larger and more fragmented.

Transformation depth also changes how earlier edits appear in the final buggy patch.
First-step added-line retention remains relatively high, dropping from 100.0\% at depth~1 to 74.9\% at depth~10.
However, first-step sequence similarity drops much more sharply, from 100.0\% to 25.9\%.
Early transformation content often remains partly present even when its original arrangement changes.
This pattern is consistent with later operators wrapping, splitting, moving, or rewriting earlier edits.
The line-accumulation factor rises from 1.00$\times$ at depth~1 to 6.63$\times$ at depth~10, indicating that intermediate patches contain substantially more changed-line content in total than the final retained patch.

Figure~\ref{fig:inject_depth_retention} further shows that later steps are more closely reflected in the final patch at depth~10.
Early steps retain 77.5\% of added lines on average, with 30.0\% sequence similarity, while late steps reach 94.9\% line retention and 80.7\% sequence similarity.
These observations are consistent with later transformations retaining parts of earlier edits while reshaping their structure.
They also show why transformation depth should not be interpreted as a simple accumulation of unchanged edits.

Together, these results provide a structural interpretation of the difficulty trend in RQ4.
Deeper transformation sequences produce larger, more fragmented patches, while later steps change the arrangement of earlier edits.
These changes may increase the code and state relationships that an agent must reason through when constructing an executable witness.
The patch metrics describe the evaluated sequences for the fixed bug type; they do not by themselves establish which structural changes cause validation failures or isolate the effect of an individual operator.

%% file: appendix/realism.tex
\section{Blinded Patch Realism Study}
\label{app:realism}

\subsection{Samples and Judgment Protocol}
We sample 700 of the 1{,}300 injected cases, stratified by repository and bug type.
Historical controls are matched by repository and edit size, also considering bug category, changed files, and hunks.
This matching aims to reduce differences in repository style and patch size that could otherwise guide source judgments.
Historical bug records establish provenance; pure refactorings, feature additions, and formatting changes are excluded.
All production-code diffs use the working-to-buggy direction, reversing historical fixes where needed.
We standardize formatting and context length, show blind IDs, and withhold commit metadata and source labels.
In-patch comments, helpers, and structure remain visible so that judgments can still reflect code-level signs of artificial edits.

Judge A is GPT-5.6-Terra with high reasoning effort; judge B is Claude Sonnet~4.6.
Both judges receive identical instructions and materials in randomized order.
Each judgment uses a fresh context, with no shared responses or source retrieval.

Each patch receives three single-patch judgments, each producing an injected probability and a brief rationale.
Probabilities of at least 0.5 yield the injected label.
Each pair receives two judgments with positions swapped, each producing a left-more-realistic probability and a rationale.
Probabilities of at least 0.5 select the left patch, and selecting the historical patch is scored as correct.
Each model makes 4{,}200 single-patch and 1{,}400 main pairwise judgments.

\paragraph{Positive Controls.}
Before judging, we sample 100 pairs and add unnecessary helpers, redundant wrappers, or unnatural comments to the injected patches, retaining their historical counterparts.
Construction checks confirm that the bug behavior is preserved.
The added artifacts contain no source labels, and controls are not selected based on judge success.
These controls test whether the judges can recognize conspicuous artificial edits under the same blinded protocol.
Evaluating both orientations adds 200 judgments per model, giving 5{,}800 judgments per model and 11{,}600 overall.
The sensitivity check requires accuracy $\geq80\%$ and a 95\% CI lower bound $>50\%$.

\subsection{Statistical Analysis}
Pairwise accuracy is the primary endpoint.
For the two correctness indicators $y_{i1},y_{i2}\in\{0,1\}$ of pair $i$, we compute $s_i=(y_{i1}+y_{i2})/2$ and average these scores over pairs.
Single-patch accuracy and AUC are auxiliary endpoints, with injected as the positive class.
Supplementary pairwise AUC uses the left-more-realistic probability and real-on-the-left as the positive class.

The equivalence hypothesis is $0.45<\theta<0.55$, against $\theta\leq0.45$ or $\theta\geq0.55$.
This study-specific margin allows deviations from chance of less than five percentage points.
Under the two one-sided testing principle ($\alpha=0.05$), each judge's complete 90\% CI must lie strictly within these bounds~\citep{lakens2017equivalence,schuirmann1987tost}.
An interval that merely includes chance is not sufficient to establish equivalence.
Approximate intervals use 10{,}000 whole-cluster bootstrap resamples~\citep{fieldwelsh2007bootstrap}, grouping patch repetitions, pair orientations, and cases sharing an original injection or historical control.
This procedure keeps related judgments together rather than treating repeated calls as independent observations.
Positive controls use 95\% CIs, and the judges are analyzed separately.

\begin{table}[H]
\centering
\footnotesize
\setlength{\tabcolsep}{3pt}
\renewcommand{\arraystretch}{1.12}
\caption{Detailed source-discrimination results. Accuracy CIs are percentages.}
\label{tab:realism_full}
\begin{tabular}{@{}llrccrc@{}}
\toprule
Task & Judge & Correct / Total & Accuracy & 90\% CI & AUC & 90\% CI \\
\midrule
Single & A & 2{,}173 / 4{,}200 & 51.74\% & 50.14--53.38 & 0.51198 & 0.49367--0.53075 \\
Single & B & 2{,}189 / 4{,}200 & 52.12\% & 50.52--53.69 & 0.51286 & 0.49426--0.53084 \\
Pairwise & A & 723 / 1{,}400 & 51.64\% & 49.07--54.21 & 0.51681 & 0.48703--0.54633 \\
Pairwise & B & 730 / 1{,}400 & 52.14\% & 49.64--54.64 & 0.52411 & 0.49517--0.55252 \\
\bottomrule
\end{tabular}
\end{table}

\begin{table}[H]
\centering
\footnotesize
\setlength{\tabcolsep}{6pt}
\renewcommand{\arraystretch}{1.1}
\caption{Artifact-bearing positive controls. Both judges meet the prespecified sensitivity check.}
\label{tab:realism_controls}
\begin{tabular}{@{}lrrccc@{}}
\toprule
Judge & Correct / Total & Incorrect / Total & Accuracy & 95\% CI & Check \\
\midrule
A & 184 / 200 & 16 / 200 & 92.00\% & 87.00\%--96.00\% & Pass \\
B & 169 / 200 & 31 / 200 & 84.50\% & 78.50\%--89.50\% & Pass \\
\bottomrule
\end{tabular}
\end{table}

Tables~\ref{tab:realism_full} and~\ref{tab:realism_controls} show that both judges meet the primary and auxiliary equivalence criteria and pass the sensitivity check.
The contrast between the main sample and positive controls suggests that the judges can detect conspicuous artifacts but have limited ability to distinguish the matched injected and historical patches.
Judge B's supplementary pairwise AUC interval reaches 0.55252, slightly above the upper equivalence bound.
Equivalence is therefore not established for this additional metric.
These findings concern source discrimination by the two evaluated judges under this protocol; they do not establish equivalence between the broader distributions of injected and historical bugs.

\subsection{Preference and Stability Diagnostics}
\begin{figure}[H]
\centering
\includegraphics[width=.85\linewidth]{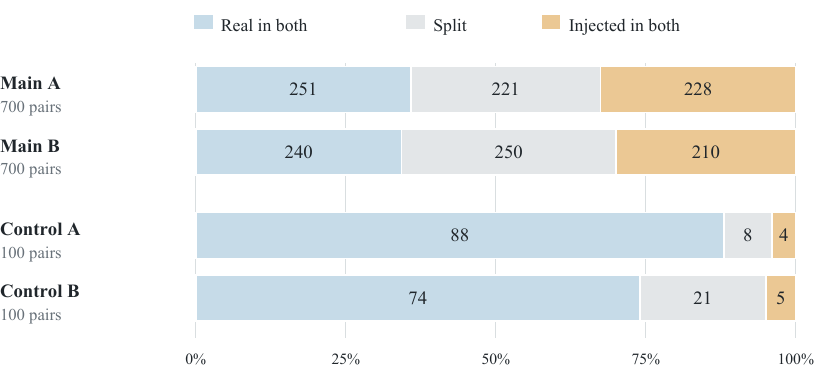}
\caption{Preferences across both orientations. Numbers give pair counts; segment widths show percentages. Main groups contain 700 pairs and control groups contain 100. ``Split'' denotes inconsistent choices, not an explicit tie.}
\label{fig:realism_preferences}
\end{figure}

In Figure~\ref{fig:realism_preferences}, \emph{split} means different underlying patch choices after swapping positions, not an explicit tie.
For real-both, split, and injected-both counts $R,S,I$, correct judgments total $2R+S$.
These totals are 723/730 for the main sample and 184/169 for controls (A/B).

\begin{table}[H]
\centering
\footnotesize
\renewcommand{\arraystretch}{1.1}
\caption{Judgment stability, position preferences, and correctness counts (1{,}400 patches per judge).}
\label{tab:realism_stability}
\begin{tabularx}{\linewidth}{@{}Xrr@{}}
\toprule
Diagnostic & Judge A & Judge B \\
\midrule
Single-patch repeat agreement & 60.62\% & 58.67\% \\
Patches with non-unanimous single judgments & 59.07\% & 62.00\% \\
Main orientation-choice agreement & 68.43\% & 64.29\% \\
Main LEFT choice rate & 52.93\% & 54.71\% \\
Main LEFT choices / judgments & 741 / 1{,}400 & 766 / 1{,}400 \\
Main split pairs choosing LEFT twice & 131 & 158 \\
\midrule
Patches with 0 / 1 correct single judgments & 277 / 369 & 247 / 402 \\
Patches with 2 / 3 correct single judgments & 458 / 296 & 466 / 285 \\
\bottomrule
\end{tabularx}
\end{table}

Repeat agreement averages the three pairwise comparisons of each patch's single-patch judgments.
Orientation agreement measures whether a judge selects the same underlying patch after positions are swapped.
More than half the patches receive non-unanimous single-patch labels, showing substantial variation across repeated judgments.
Table~\ref{tab:realism_stability} also shows a left-choice tendency, especially for judge B.
Evaluating both orientations balances the position of the historical patch, but does not establish position invariance.
The aggregate equivalence results should therefore not be interpreted as stable decisions on every individual patch.

For the per-patch counts $N_0,\ldots,N_3$ in Table~\ref{tab:realism_stability}, correct judgments equal $N_1+2N_2+3N_3$.
Repeat agreement equals $[3(N_0+N_3)+N_1+N_2]/4{,}200$, since unanimous labels agree in all three comparisons and non-unanimous labels agree in only one.

%% file: appendix/prompt.tex
\section{Prompt Design}
\label{appendix:prompt}

We report the prompt structures used in \toolname{}.
The implementation uses layered prompts: a reusable system prompt specifies global behavioral constraints, while a dynamically assembled task payload injects case-specific fields such as the bug type, bug type specification, transformation operator, execution context, replay command, candidate anchors, and file scope.

For readability, repository-local paths and long execution context lists are shortened in this appendix, but the task goals, constraints, input fields, and required output formats follow the actual pipeline.

\providecolor{backcolour}{HTML}{FAFAFA}
\providecolor{promptback}{HTML}{FAFAFA}

\newtcolorbox{pocpromptbox}[1]{
  title={#1},
  width=\linewidth,
  colback=white,
  colframe=gray!60,
  colbacktitle=gray!60,
  coltitle=white,
  fonttitle=\bfseries,
  fontupper=\footnotesize\sloppy\raggedright,
  boxrule=1pt,
  arc=2mm,
  left=5pt,
  right=5pt,
  top=4pt,
  bottom=4pt,
  before skip=4pt,
  after skip=5pt,
  breakable
}

\newtcblisting{pocpromptlisting}[1]{
  title={#1},
  width=\linewidth,
  listing only,
  listing engine=listings,
  enhanced,
  breakable,
  colback=white,
  colframe=gray!60,
  colbacktitle=gray!60,
  coltitle=white,
  fonttitle=\bfseries,
  boxrule=1pt,
  arc=2mm,
  left=5pt,
  right=5pt,
  top=4pt,
  bottom=4pt,
  before skip=4pt,
  after skip=5pt,
  listing options={
    language={},
    basicstyle=\ttfamily\scriptsize,
    keywordstyle=\ttfamily\scriptsize,
    ndkeywordstyle=\ttfamily\scriptsize,
    identifierstyle=\ttfamily\scriptsize,
    stringstyle=\ttfamily\scriptsize,
    commentstyle=\ttfamily\scriptsize,
    emphstyle=\ttfamily\scriptsize,
    backgroundcolor=\color{promptback},
    breaklines=true,
    breakatwhitespace=false,
    columns=fullflexible,
    keepspaces=true,
    showstringspaces=false,
    showtabs=false,
    tabsize=2
  }
}

\suppressfloats[t]
\begin{table}[!htbp]
\centering
\footnotesize
\renewcommand{\arraystretch}{1.04}
\setlength{\tabcolsep}{4pt}
\caption{Layered prompt structures used by \toolname{} during benchmark construction and executable bug validation.}
\label{tab:prompt_templates}
\begin{tabularx}{\textwidth}{p{0.29\textwidth}XX}
\toprule
\textbf{Prompt Structure} & \textbf{Main Inputs} & \textbf{Required Output} \\
\midrule
Bug Injection System Prompt
& Global injection rules, replay-command discipline, bounded edit--run--revise loop, diff checks, and success criteria
& Single-line \texttt{INJECT\_REPORT\_JSON} with operational evidence fields \\
Bug Injection Task Payload
& Bug type, bug type specification, required and forbidden elements, execution context, candidate anchors, trigger hints, and file scope
& Same \texttt{INJECT\_REPORT\_JSON} prefix with type-site rationale fields \\
Bug-Preserving Transformation Task Payload
& Transformation operator specification, applicability requirements, preservation constraints, execution context, and candidate anchors
& Single-line \texttt{TRANSFORM\_REPORT\_JSON} \\
Bug Validation System Prompt
& Test-only edit constraints, Java test-case materialization rules, executable-witness objective, and evaluation-attempt policy
& Single-line \texttt{TRIGGER\_REPORT\_JSON} fallback schema \\
Bug Validation Task Payload
& Buggy repository metadata, case ID, bug type, applied transformations, bug-related files, replay command, and optional execution context
& Single-line \texttt{TRIGGER\_REPORT\_JSON} with detailed witness-construction fields \\
\bottomrule
\end{tabularx}
\end{table}

\subsection{Bug Injection System Prompt}
\label{appendix:bug_injection_system_prompt}

\begin{pocpromptbox}{Bug Injection System Prompt}
\textbf{Mission.}
Introduce a real, localized bug into production code so that the externally provided replay command fails for the intended bug type.
The repository must end with a non-empty semantic git diff, and the replay command must fail because of the injected bug.

\smallskip
\textbf{Core rules.}
\begin{enumerate}[leftmargin=*,topsep=2pt,itemsep=1pt,parsep=0pt,partopsep=0pt]
  \item Produce a non-empty semantic production-code diff.
  \item Run the replay command exactly as provided, without changing the module, flags, test target, or command structure.
  \item Do not modify tests unless explicitly allowed by the task.
  \item Do not modify build files, dependency files, CI files, or repository configuration.
  \item Do not change public or protected API signatures.
  \item Keep the edit localized, realistic, and consistent with the repository style.
  \item Do not claim success if the git diff is empty or if the observed failure is unrelated to the injected bug.
\end{enumerate}

\smallskip
\textbf{Construction-side execution budget.}
During one bug-injection session, the implementation permits at most three runs of the provided replay command.
If the replay command still passes, the agent must revise the production-code edit and rerun the same command within this budget.
This construction-side replay budget is separate from the bug-validation attempt budget used in the evaluation.

\smallskip
\textbf{Required final output.}
The system prompt requires exactly one final single-line JSON object prefixed by \texttt{INJECT\_REPORT\_JSON}.
The operational evidence fields include:
\texttt{ok}, \texttt{git\_changed\_files}, \texttt{repro\_cmd}, \texttt{repro\_exit\_code}, \texttt{observed\_exception}, \texttt{call\_chain}, and \texttt{notes}.
\end{pocpromptbox}

\subsection{Bug Injection Task Payload}
\label{appendix:bug_injection_task_payload}

\begin{pocpromptbox}{Bug Injection Task Payload}
\textbf{Task-specific inputs.}
\begin{itemize}[leftmargin=*,topsep=2pt,itemsep=1pt,parsep=0pt,partopsep=0pt]
  \item Bug type and bug type ID.
  \item Bug type name, category, scope, and language.
  \item Bug type description.
  \item Required elements that must appear in the patch.
  \item Forbidden elements that must not appear in the patch.
  \item Expected failure signal and trigger-condition hints.
  \item Execution context summary collected from the target test.
  \item Candidate anchors with trace-relevant methods, files, and candidate edit sites.
  \item Allowed or preferred production-code file scope.
\end{itemize}

\smallskip
\textbf{Bug type contract.}
The injected bug must match the provided bug type specification rather than merely resemble it.
Required elements must be visible in the actual patch, and forbidden elements must not appear in the final patch.
The chosen site should be trace-relevant and should allow the provided replay command to trigger the injected bug after validation.

\smallskip
\textbf{Task-level reporting fields.}
The task payload uses the same \texttt{INJECT\_REPORT\_JSON} prefix and adds bug-type-site rationale fields:
\texttt{ok}, \texttt{type\_id}, \texttt{target\_files}, \texttt{target\_symbols}, \texttt{selected\_anchor}, \texttt{why\_this\_site}, \texttt{expected\_trigger\_path}, \texttt{stealth\_rationale}, \texttt{bug\_mechanism}, \texttt{preserved\_prior\_structure}, and \texttt{possible\_risks}.
The implementation field \texttt{stealth\_rationale} records the localization and plausibility of the edit.
\end{pocpromptbox}

\subsection{Bug-Preserving Transformation Task Payload}
\label{appendix:bug_preserving_transformation_task_payload}

\begin{pocpromptbox}{Bug-Preserving Transformation Task Payload}
The transformation prompt is a dynamically assembled task payload generated from an internal transformation operator specification.
The payload uses implementation-level fields such as \texttt{id}, \texttt{name}, \texttt{intent}, \texttt{description}, \texttt{applicability}, \texttt{forbidden\_elements}, \texttt{must\_preserve}, and \texttt{post\_conditions}.
Each internal operator is mapped to a paper-facing transformation operator and family.
For example, \texttt{t\_rare\_profile\_gate\_v1} is reported as Rare Input Path Conditioning, \texttt{t\_call\_stack\_deepen\_v1} is reported as Helper Call Insertion, and \texttt{t\_private\_adapter\_layer\_v1} is reported as a call-site rerouting / private-adapter transformation.

\smallskip
\textbf{Task.}
Apply one bug-preserving transformation step after successful bug injection.
The transformation should alter the surrounding production-code structure while preserving compilation and preserving the replay behavior that exposes the injected bug.
It is bug-preserving in the benchmark-construction sense: it preserves the relevant trigger path and bug-inducing behavior, but it is not required to preserve the full behavior of the original unmodified program.

\smallskip
\textbf{Inputs.}
\begin{itemize}[leftmargin=*,topsep=2pt,itemsep=1pt,parsep=0pt,partopsep=0pt]
  \item Internal transformation ID and operator name.
  \item Paper-facing operator term and transformation family.
  \item Operator intent and description.
  \item Applicability requirements.
  \item Forbidden edits.
  \item Must-preserve constraints, including compilation success, public API signatures, behavior for non-trigger inputs when applicable, and replay-command triggerability.
  \item Post-conditions expected from the operator.
  \item Execution context summary and candidate anchors.
\end{itemize}

\smallskip
\textbf{Edit boundaries.}
\begin{enumerate}[leftmargin=*,topsep=2pt,itemsep=1pt,parsep=0pt,partopsep=0pt]
  \item Do not modify tests unless explicitly allowed by the operator.
  \item Do not change public or protected API signatures.
  \item Do not perform broad refactors, renames, or formatting sweeps.
  \item Keep the transformation localized to trace-relevant production code.
  \item Preserve the injected bug-triggering behavior under the same replay command.
\end{enumerate}

\smallskip
\textbf{Required final output.}
The payload requires exactly one final single-line JSON object prefixed by \texttt{TRANSFORM\_REPORT\_JSON}.
The JSON object contains:
\texttt{ok}, \texttt{transform\_id}, \texttt{changed\_files}, \texttt{edit\_sites}, \texttt{what\_was\_added\_or\_wrapped}, \texttt{what\_was\_preserved}, \texttt{what\_was\_replaced\_or\_overwritten}, \texttt{compatibility\_with\_previous\_steps}, \texttt{expected\_effect\_on\_stealth}, and \texttt{expected\_effect\_on\_trigger\_depth}.
The field \texttt{expected\_effect\_on\_stealth} is the implementation field name for the expected surface-form and review-plausibility effect of the transformation.
\end{pocpromptbox}

\subsection{Bug Validation System Prompt}
\label{appendix:poc_generation_system_prompt}

\begin{pocpromptbox}{Bug Validation System Prompt}
\textbf{Mission.}
Create a Java test that exposes an already-injected production bug under the provided Maven replay command.
The agent's job is not to fix code, improve the repository, or write a generic regression test.
The only goal is to materialize a test that exposes the injected bug.

\smallskip
\textbf{Evaluation attempt budget.}
For each benchmark case, the evaluation protocol allows at most three bug-validation attempts.
Each attempt must follow the same test-only constraints and use the provided replay command as the validation target.
A run succeeds only if the generated witness test compiles, the replay command exposes the injected bug, and no production file or build configuration is modified.

\smallskip
\textbf{Hard constraints.}
\begin{enumerate}[leftmargin=*,topsep=2pt,itemsep=1pt,parsep=0pt,partopsep=0pt]
  \item Never modify production code.
  \item Never modify build files, dependency files, CI files, configuration files, or repository metadata.
  \item Only create or edit Java test files under \texttt{src/test/java}.
  \item Prefer rewriting the official test file at the provided path and with the provided package and class name.
  \item The witness test must materialize in the repository as a real \texttt{.java} file before verification.
  \item The test must be designed to trigger the injected bug, not to validate normal behavior.
  \item Do not delete or modify unrelated files.
\end{enumerate}

\smallskip
\textbf{Required final output.}
The system prompt requires a final single-line \texttt{TRIGGER\_REPORT\_JSON}.
The fallback schema includes:
\texttt{ok}, \texttt{test\_strategy}, \texttt{target\_bug\_path\_hypothesis}, \texttt{why\_trigger\_should\_fail}, and \texttt{materialized\_test\_files}.
\end{pocpromptbox}

\subsection{Bug Validation Task Payload}
\label{appendix:poc_generation_task_payload}

\begin{pocpromptbox}{Bug Validation Task Payload}
\textbf{Task-specific inputs.}
\begin{itemize}[leftmargin=*,topsep=2pt,itemsep=1pt,parsep=0pt,partopsep=0pt]
  \item Repository name.
  \item Evaluation group or context condition.
  \item Injected case run ID.
  \item Bug type ID and bug type.
  \item Applied bug-preserving transformations.
  \item Bug-related production files and location metadata.
  \item Injection-stage verification exit code.
  \item Maven replay command used for validation.
  \item Natural-language bug summary.
  \item Target test path and target test class when available.
\end{itemize}

\smallskip
\textbf{Execution context handling.}
The two evaluation settings are \(\mathsf{NoEC}\) and \(\mathsf{WithEC}\).
In \(\mathsf{WithEC}\), the task payload includes the execution context summary collected during benchmark construction.
In \(\mathsf{NoEC}\), this execution context block is omitted.
Neither setting provides the construction-time witness, hidden validation oracle, construction-time solution code, or permission to modify production code.

\smallskip
\textbf{Required final output.}
The task payload asks for a detailed final \texttt{TRIGGER\_REPORT\_JSON} schema with:
\texttt{summary}, \texttt{suspected\_bug\_mechanism}, \texttt{inspected\_production\_files}, \texttt{inspected\_test\_files}, \texttt{selected\_test\_path}, \texttt{selected\_test\_class}, \texttt{trigger\_strategy}, \texttt{expected\_failure\_signal}, \texttt{materialized\_test\_files}, and \texttt{notes}.
\end{pocpromptbox}

\subsection{End-to-End Case Prompt Example}
\label{appendix:end_to_end_prompt_example}

This example shows a concrete case-level prompt flow assembled by the pipeline.
The case uses repository \texttt{dubbo}, bug type Guard Predicate Inversion, and a depth-3 transformation chain.
Long execution context lists are shortened for page fit, but the fields and constraints shown below follow the actual prompt structure.

\begin{pocpromptlisting}{Case Metadata}
run_id: inject__20260315T064448Z
repository: dubbo
trace_group: short
trace_id: org.apache.dubbo.common.url.URLParamTest
bug_type: Guard Predicate Inversion
type_id: logic_predicate_inversion_in_guard
transform_depth: 3
transform_ids:
  - t_rare_profile_gate_v1
  - t_call_stack_deepen_v1
  - t_private_adapter_layer_v1
paper_facing_transform_terms:
  - Rare Input Path Conditioning
  - Helper Call Insertion
  - Call-Site Rerouting / Private Adapter Layer
official_test_path: dubbo-common/src/test/java/org/apache/dubbo/common/url/URLParamTest.java
official_test_class: org.apache.dubbo.common.url.URLParamTest
replay_command: ./mvnw -q -pl dubbo-common -Dtest=org.apache.dubbo.common.url.URLParamTest test
context_setting_for_executable_validation: WithEC
evaluation_attempt_budget: at most 3 bug-validation attempts per benchmark case
\end{pocpromptlisting}

\begin{pocpromptlisting}{Bug Injection Prompt}
SYSTEM:
You are operating inside a Git repository in an automated bug-injection pipeline.

MISSION:
Introduce a real, localized bug into production code so that the externally provided reproduction command fails for the intended bug type. Your work is only valid if the repository ends the session with a non-empty semantic git diff and the provided reproduction command, run verbatim, fails because of the injected bug.

NON-NEGOTIABLE RULES:
- You must produce a non-empty semantic git diff in the current repository state before finalizing.
- Semantic git diff means a real logic-changing edit, not whitespace-only, formatting-only, comment-only, import-only, rename-only, or mechanically trivial changes.
- You must run the reproduction command exactly as provided in the task, verbatim.
- Do not invent, replace, shorten, simplify, or substitute any command, module, flag, path, or test target.
- Do not modify test code unless the task explicitly allows it.
- Do not change public or protected API signatures.
- Do not claim success based on a hypothetical edit.
- Do not output a final success JSON if git diff is empty.
- Do not output a final success JSON if the reproduction command in your JSON differs from the externally provided command.

CONSTRUCTION-SIDE EXECUTION BUDGET:
- Up to 3 total runs of the provided reproduction command in the same injection session.
- If the reproduction command passes, make another production-code edit and rerun the same command verbatim within this budget.

TASK PAYLOAD:
You are performing controlled BUG INJECTION into production code for research benchmarking.
Goal: introduce a subtle, realistic bug consistent with the specified bug type, while keeping changes localized and review-plausible.

TYPE:
- id: logic_predicate_inversion_in_guard
- name: Predicate inversion in critical guard
- category: Logic
- scope: intra_method_branch_condition
- language: java

Description:
A boolean predicate used to guard important logic is inverted or altered in a subtle way. The code still compiles, branches look plausible, but the semantics flip which path handles which inputs, causing assertions and invariants to fail.

Required elements:
- if_or_else_block_guarding_non_trivial_logic
- condition_based_on_collection_size_or_status_flag
- tests_that_distinguish_between_valid_and_invalid_input_cases

Forbidden elements:
- conditions_that_are_constant_true_or_false_after_modification
- completely_removing_the_guard_and_inlining_the_body
- branches_with_no_observable_effect

Runtime effect hints:
- exception_type_hint: none
- typical_failure_signal: assertion failure or invalid return value

Trigger condition hints:
- test_shape_hint: tests feed different valid/invalid input categories and assert different behavior
- input_properties_hint: at least one test relies on the original guard semantics

TRACE CONTEXT SUMMARY:
{
  "trace_id": "org.apache.dubbo.common.url.URLParamTest",
  "test_name": "org.apache.dubbo.common.url.URLParamTest",
  "methods_count": 91
}

CANDIDATE ANCHORS:
- method: URLParamTest
  file: dubbo-common/src/test/java/org/apache/dubbo/common/url/URLParamTest.java
  role: target test
- method: hasMethodParameter
  file: dubbo-common/src/main/java/org/apache/dubbo/common/url/component/URLParam.java
  role: trace-relevant guard candidate
- method: getMethodParameter
  file: dubbo-common/src/main/java/org/apache/dubbo/common/url/component/URLParam.java
  role: trace-relevant lookup candidate
- method: getMethodParameterStrict
  file: dubbo-common/src/main/java/org/apache/dubbo/common/url/component/URLParam.java
  role: trace-relevant lookup candidate
- method: initMethodParameters
  file: dubbo-common/src/main/java/org/apache/dubbo/common/url/component/URLParam.java
  role: method-parameter initialization path
- method: size
  file: dubbo-common/src/main/java/org/apache/dubbo/common/url/component/URLParam.java
  role: status/collection-size related helper
- method: isEmpty
  file: dubbo-common/src/main/java/org/apache/dubbo/common/url/component/URLParam.java
  role: status/collection-size related helper

DELIVERABLE:
1. Implement the bug strictly according to the bug type contract.
2. Keep the patch minimal and consistent with repository style.
3. Run the provided reproduction command verbatim.
4. Provide evidence in the final JSON.

REQUIRED FINAL OUTPUT:
Append exactly one final line with this prefix:
INJECT_REPORT_JSON: { ... }

Required field groups:
- operational evidence: ok, git_changed_files, repro_cmd, repro_exit_code, observed_exception, call_chain, notes
- type-site rationale: type_id, target_files, target_symbols, selected_anchor, why_this_site, expected_trigger_path, stealth_rationale, bug_mechanism, preserved_prior_structure, possible_risks
\end{pocpromptlisting}

\begin{pocpromptlisting}{Bug Injection Report}
INJECT_REPORT_JSON:
{
  "ok": true,
  "type_id": "logic_predicate_inversion_in_guard",
  "target_files": [
    "dubbo-common/src/main/java/org/apache/dubbo/common/url/component/URLParam.java"
  ],
  "target_symbols": [
    "getAnyMethodParameter"
  ],
  "selected_anchor": "getAnyMethodParameter",
  "why_this_site": "The method is reached by URLParamTest and contains a critical guard over method-specific parameter lookup.",
  "expected_trigger_path": "URLParamTest -> URLParam.getAnyMethodParameter -> inverted guard returns null instead of the expected method parameter.",
  "stealth_rationale": "The edit is a localized predicate substitution inside an existing defensive check.",
  "bug_mechanism": "A guard condition is inverted from a non-empty map check to an empty map check, swapping valid and invalid lookup paths.",
  "preserved_prior_structure": "Method signatures, surrounding branch structure, and other production logic remain unchanged.",
  "possible_risks": "The bug is most visible through method-parameter tests that expect lookup results from non-empty maps."
}
\end{pocpromptlisting}

\begin{pocpromptlisting}{Bug-Preserving Transformation Prompt}
TASK:
Apply one bug-preserving transformation step after successful bug injection.
The transformation must keep the repository compilable and preserve the replay behavior that exposes the injected bug.

TRANSFORM SPEC 1:
- id: t_rare_profile_gate_v1
- paper_facing_operator: Rare Input Path Conditioning
- family: Control Flow
- name: Gate misroute behind rare profile
- description: restrict the misroute to a deterministic but low-probability input profile
- applicability: at_least_one_mid_level_helper_candidate
- forbidden_edits: changes_to_public_API_signatures, obvious_dead_branches_such_as_if_false
- must_preserve: compilation_success, public_api_signatures, replay_command_triggerability
- post_conditions: rare deterministic gate on the bug-related path

TRANSFORM SPEC 2:
- id: t_call_stack_deepen_v1
- paper_facing_operator: Helper Call Insertion
- family: Call Stack
- name: Deepen call stack with a natural helper extraction
- description: introduce a small helper method or lightweight wrapper to deepen the call stack
- applicability: at_least_one_mid_level_helper_candidate
- forbidden_edits: changes_to_public_API_signatures, broad refactors, formatting sweeps
- must_preserve: compilation_success, public_api_signatures, behavior_for_non_trigger_inputs
- post_conditions: additional_helper_in_call_chain

TRANSFORM SPEC 3:
- id: t_private_adapter_layer_v1
- paper_facing_operator: Call-Site Rerouting / Private Adapter Layer
- family: Call Stack
- name: Add a private adapter layer and reroute one call site
- description: introduce a private adapter method and reroute at least one internal call site through it
- applicability: at_least_one_internal_call_site_can_be_rerouted
- forbidden_edits: changes_to_public_API_signatures, obvious_dead_branches_such_as_if_false
- must_preserve: compilation_success, public_api_signatures, behavior_for_non_trigger_inputs
- post_conditions: additional_adapter_in_call_chain

TRACE CONTEXT SUMMARY:
{
  "trace_id": "org.apache.dubbo.common.url.URLParamTest",
  "test_name": "org.apache.dubbo.common.url.URLParamTest",
  "methods_count": 91
}

CANDIDATE ANCHORS:
- URLParam.hasMethodParameter
- URLParam.getMethodParameter
- URLParam.getMethodParameterStrict
- URLParam.initMethodParameters
- URLParam.size
- URLParam.isEmpty

EDIT BOUNDARIES:
- Do not modify tests unless explicitly allowed by the transform specification.
- Do not change public or protected API signatures.
- Avoid broad refactors, renames, or formatting sweeps.
- Keep diffs localized to trace-relevant production code.
- Preserve replay-command triggerability.

REQUIRED FINAL OUTPUT:
Append exactly one final line with this prefix:
TRANSFORM_REPORT_JSON: { ... }

Required keys:
ok, transform_id, changed_files, edit_sites, what_was_added_or_wrapped, what_was_preserved, what_was_replaced_or_overwritten, compatibility_with_previous_steps, expected_effect_on_stealth, expected_effect_on_trigger_depth
\end{pocpromptlisting}

\begin{pocpromptlisting}{Bug-Preserving Transformation Report}
TRANSFORM_REPORT_JSON:
{
  "ok": true,
  "transform_id": "t_rare_profile_gate_v1 + t_call_stack_deepen_v1 + t_private_adapter_layer_v1",
  "changed_files": [
    "dubbo-common/src/main/java/org/apache/dubbo/common/url/component/URLParam.java"
  ],
  "edit_sites": [
    "URLParam.getMethodParameter",
    "URLParam.hasMethodParameter",
    "URLParam.keyIndexToOffset",
    "URLParam.computeValueOffset",
    "URLParam.resolveParameterWithEncoding"
  ],
  "what_was_added_or_wrapped": "Added private helper and adapter methods around parameter lookup and offset resolution.",
  "what_was_preserved": "Public APIs, compilation, and replay-command triggerability.",
  "what_was_replaced_or_overwritten": "Direct parameter and offset resolution call sites were rerouted through private helpers.",
  "compatibility_with_previous_steps": "The transformation chain preserves the injected guard-inversion behavior.",
  "expected_effect_on_stealth": "The resulting structure resembles localized helper extraction and profile-aware lookup logic.",
  "expected_effect_on_trigger_depth": "The execution path now includes additional helper and adapter calls before the wrong value is observed."
}
\end{pocpromptlisting}

\begin{pocpromptlisting}{Bug Validation Prompt}
SYSTEM:
You are an autonomous code agent operating inside a Java repository.

You are running an executable bug-validation task for an already-injected production bug.
Your job is not to fix code.
Your job is not to improve the repository.
Your job is not to write a generally reasonable test.
Your only goal is to create a new Java test case that exposes the already-injected bug under the provided Maven verification command.

EVALUATION ATTEMPT BUDGET:
For each benchmark case, the evaluation protocol allows at most three bug-validation attempts.
Each attempt must obey the same test-only constraints and use the provided replay command as the validation target.

HARD CONSTRAINTS:
1. Never modify production code under src/main/java or any non-test source root.
2. Never modify build files, pom.xml files, Gradle files, CI files, config files, or dependency files.
3. Only create or edit Java test files under src/test/java.
4. Prefer rewriting the official testcase file at the same path and with the same package and class name when that information is provided.
5. The testcase must materialize in the repository as a real .java file before verification.
6. The testcase must be designed to trigger the injected bug, not to validate normal behavior.
7. Do not output file blocks for an external parser. Directly write files into the repository workspace.
8. Keep the testcase minimal but bug-targeted.
9. Do not delete or modify unrelated files.

TASK PAYLOAD:
Generate a new Java testcase that triggers the already-injected bug.

Success criteria:
- A real Java testcase file exists in the repository under src/test/java.
- The testcase uses the correct package declaration.
- The testcase compiles.
- The provided verification command fails because the injected bug is triggered.
- The failure is a semantic test failure, not a compile error and not a harness failure.
- No production file or build configuration file is modified.

Repository: dubbo
Context setting: WithEC
Implementation group: B
Injected case run_id: inject__20260315T064448Z
Type ID: logic_predicate_inversion_in_guard
Bug type: Guard Predicate Inversion
Applied transforms: t_rare_profile_gate_v1, t_call_stack_deepen_v1, t_private_adapter_layer_v1
Bug-related production files:
- dubbo-common/src/main/java/org/apache/dubbo/common/url/component/URLParam.java
Official test path: dubbo-common/src/test/java/org/apache/dubbo/common/url/URLParamTest.java
Official test class: org.apache.dubbo.common.url.URLParamTest
Verification command: ./mvnw -q -pl dubbo-common -Dtest=org.apache.dubbo.common.url.URLParamTest test
Injection-stage verify rc: 1

Type summary:
A boolean guard condition in production code was inverted, causing guarded URL-parameter lookup logic to execute under the wrong condition and leading to semantically incorrect behavior without changing public APIs.

Execution context block:
The target test reaches URLParam construction and method-specific parameter lookup.
Trace-relevant production methods include URLParam.hasMethodParameter, URLParam.getMethodParameter, URLParam.getMethodParameterStrict, URLParam.initMethodParameters, URLParam.size, and URLParam.isEmpty.
This execution context block identifies the path used during benchmark construction; it does not include the reference-trigger implementation or hidden validation oracle.

What you must do:
- Read the bug context carefully.
- Inspect the relevant production and test files.
- Rewrite or create the testcase directly at the official test path.
- Keep the package name and class name correct.
- Exercise the suspected buggy path through URLParam method-specific parameter lookup.
- Write an oracle that fails under the injected bug.
- Run the provided verification command if needed.
- Leave no production-code or build-configuration changes.

Required final output:
At the very end, print exactly one single-line JSON object prefixed by:
TRIGGER_REPORT_JSON:

The JSON object must have this schema:
{
  "summary": "...",
  "suspected_bug_mechanism": "...",
  "inspected_production_files": ["..."],
  "inspected_test_files": ["..."],
  "selected_test_path": "...",
  "selected_test_class": "...",
  "trigger_strategy": ["...", "...", "..."],
  "expected_failure_signal": "...",
  "materialized_test_files": ["..."],
  "notes": "..."
}
\end{pocpromptlisting}

\begin{pocpromptlisting}{Bug Validation Report}
TRIGGER_REPORT_JSON:
{
  "summary": "Generate a URLParam test that exercises method-specific parameter lookup after the transformed guard inversion.",
  "suspected_bug_mechanism": "A guard inversion and transformed lookup path cause method-specific parameter retrieval to return an incorrect value or miss a valid method entry.",
  "inspected_production_files": [
    "dubbo-common/src/main/java/org/apache/dubbo/common/url/component/URLParam.java"
  ],
  "inspected_test_files": [
    "dubbo-common/src/test/java/org/apache/dubbo/common/url/URLParamTest.java"
  ],
  "selected_test_path": "dubbo-common/src/test/java/org/apache/dubbo/common/url/URLParamTest.java",
  "selected_test_class": "org.apache.dubbo.common.url.URLParamTest",
  "trigger_strategy": [
    "Construct URLParam with multiple parameters so that method-specific lookup and transformed helper paths are exercised.",
    "Call hasMethodParameter and getMethodParameter on methods that should be present.",
    "Assert the expected valid lookup result so that the injected guard inversion becomes observable."
  ],
  "expected_failure_signal": "Assertion failure: expected method-specific parameter to be present or equal to the expected value, but the transformed buggy path returns an incorrect result.",
  "materialized_test_files": [
    "dubbo-common/src/test/java/org/apache/dubbo/common/url/URLParamTest.java"
  ],
  "notes": "The test is written only under src/test/java and does not modify production code or build configuration."
}
\end{pocpromptlisting}